\documentstyle[preprint,eqsecnum,aps]{revtex}

\begin{document}
\draft

\title{Femtometer Toroidal Structures in Nuclei}

\author{J. L. Forest\cite{jlf} and V. R. Pandharipande\cite{vrp}}
\address{Department of Physics, University of Illinois at Urbana-Champaign,
Urbana, IL 61801}
\author{Steven C. Pieper\cite{scp} and R. B. Wiringa\cite{rbw}}
\address{Physics Division, Argonne National Laboratory, Argonne, IL 60439}
\author{R. Schiavilla\cite{rs}}
\address{CEBAF Theory Group, Newport News, VA 23606, \\
and \\
Department of Physics, Old Dominion University, Norfolk, VA 23529}
\author{A. Arriaga\cite{aa}}
\address{Centro de Fisica Nuclear da Universidade de Lisboa,
         Avenida Gama Pinto 2, 1699 Lisboa, Portugal}

\date{\today}
\maketitle
\newpage
\begin{abstract}
The two-nucleon density distributions in states with isospin $T=0$, spin
$S$=1 and projection $M_S$=0 and $\pm$1 are studied in $^2$H, $^{3,4}$He,
$^{6,7}$Li and $^{16}$O. The equidensity surfaces for $M_S$=0 distributions
are found to be toroidal in shape, while those of $M_S$=$\pm$1 have dumbbell
shapes at large density. The dumbbell shapes are generated by rotating
tori. The toroidal shapes indicate that the tensor correlations have near
maximal strength at $r<2$ fm in all these nuclei. They provide new insights
and simple explanations of the structure and electromagnetic form factors
of the deuteron, the quasi-deuteron model, and the $dp$, $dd$ and
$\alpha d$ $L$=2 ($D$-wave) components in $^3$He, $^4$He and $^6$Li.
The toroidal distribution has a maximum-density diameter of $\sim$1 fm
and a half-maximum density thickness of $\sim$0.9 fm.  Many realistic
models of nuclear forces predict these values, which are supported by
the observed
electromagnetic form factors of the deuteron, and also predicted by classical
Skyrme effective Lagrangians, related to QCD in the limit of infinite colors.
Due to the rather small size of this structure, it could have a revealing
relation to certain aspects of QCD. Experiments to probe this structure
and its effects in nuclei are suggested. Pair distribution functions in
other $T,S$ channels are also discussed; those in $T,S = 1,1$ have
anisotropies expected from one-pion exchange interactions. The tensor
correlations in $T,S = 0,1$ states are found to deplete the number of
$T,S = 1,0$ pairs in nuclei and cause a reduction in nuclear binding
energies via many-body effects.
\end{abstract}

\newpage

\section{INTRODUCTION}

Nuclear structure has been discussed mostly in the context of the liquid
drop and shell models. These models have been extremely useful in explaining
many observed nuclear properties. However, they are based on
macroscopic concepts, and do not address the simplest nuclei,
hydrogen and helium.  Furthermore, recent $(e,e^{\prime}p)$
experiments~\cite{ref1} have indicated that in heavier near-closed-shell
nuclei, less than $70\%$ of the nucleons are in the single-particle
orbitals that would be fully occupied in the simple shell model.

To obtain a more microscopic description of nuclear structure,
we may regard the nucleus as a collection of interacting nucleons described
by the Hamiltonian~\cite{ref2}
\begin{equation}
H=\sum_i-\frac{\hbar^2}{2m}\nabla_i^2+\sum_{i<j}v_{ij}+\sum_{i<j<k}V_{ijk} .
\end{equation}
The interactions
$v_{ij}$ and $V_{ijk}$ are not exactly known, but $v_{ij}$ is well constrained
by the available scattering data, and binding energies and theoretical
considerations place important constraints on $V_{ijk}$.
The structure of the ground-state wave function $\Psi_0$
at small interparticle distances is influenced by the repulsive core and
tensor parts of $v_{ij}$. Most realistic models of $v_{ij}$ contain
these components and, for example, the Reid~\cite{Rei68}, Paris~\cite{Lac80},
Urbana~\cite{Lag81} and new Argonne $v_{18}$~\cite{Wir95} models seem to
predict
similar structures. The three-nucleon interaction $V_{ijk}$ is much weaker
than the $v_{ij}$. Due to a large cancellation between kinetic and two-body
interaction energies, it has a significant effect on nuclear binding
energies~\cite{ref2} but its effect on the structure of $\Psi_0$ is much less
than that of the better known $v_{ij}$.

Due to the strong spin-isospin dependence of $v_{ij}$ and $V_{ijk}$ it is
difficult to solve the Schr\"{o}dinger equation with the Hamiltonian (1.1).
Only recently it has been possible to obtain accurate solutions for
$A\leq7$ nuclei~\cite{PPCW95,pudliner} with the Green's Function Monte Carlo
(GFMC) method. Accurate variational wave functions $\Psi_v$, which contain
less than $0.5\%$ admixture of excited states are known for $A$ = 3 and 4
nuclei~\cite{APW95}.  The available $\Psi_v$ for $^{6,7}$Li~\cite{WPP96} and
$^{16}$O~\cite{PWP92,Pie96} are certainly not as accurate as those for
$A$ = 3 and 4, nevertheless they presumably contain most of the important
structure of the $\Psi_0$.

In this paper we examine the short-range structure of $^2$H, $^{3,4}$He,
$^{6,7}$Li and $^{16}$O by calculating the two-nucleon density distributions
in states with isospin $T$, spin $S$, and spin projection $M_S$. Since the
deuteron has only two nucleons, its one- and two-body density distributions
are trivially related. Variational wave functions $\Psi_v$ and Monte Carlo
methods are used for $A > 2$.

The two-nucleon distributions in the $T,S = 0,1$ states have a strong
dependence on the spin projection $M_S$. The equidensity surfaces, spanning
the top three quarters of the density range in $M_S$=0 states, have toroidal
shape. These tori are produced by the joint action of the repulsive core
and tensor interactions. In contrast the equidensity shapes in the
$M_S$=$\pm$1 have dumbbell shapes, which have been studied earlier in the
deuteron~\cite{ericson,Blatt}. A brief description of the two-nucleon
interaction
in $T,S = 0,1$ states is given in section II, and the density distribution
of the deuteron is discussed in detail in section III, where we show that the
dumbbell-shaped distributions in $M_S$=$\pm$1 states are produced by
rotating tori. Commonly used models of $v_{ij}$ predict that the maximum
density torus has a diameter of $\sim$1 fm, and the half-maximum density
torus has a thickness of $\sim$0.9 fm. In section III we relate these
dimensions of the toroidal distribution to the observed electromagnetic
form factors of the deuteron. The structures are rather dense; current
models predict the maximum one-body density of the torus inside the
deuteron to be $\sim 0.34\pm 0.02$ fm$^{-3}$, i.e., approximately twice
nuclear matter density.

The two-nucleon $T,S = 0,1$, $M_S = 0,\pm1$ density distributions in
$^{3,4}$He, $^{6,7}$Li and $^{16}$O are compared with those of the deuteron
in section IV. The distributions for $r<2$ fm differ only by a single scale
factor. They indicate that in the $T=0$ state, the tensor correlations
have near maximal strength for $r<2$ fm in all these nuclei. The scale
factor is identified as the Levinger-Bethe quasi-deuteron number, and
its value is compared to the ratios of total photon ($E_{\gamma}$ = 80 to
120 MeV) and pion ($E_{\pi^+}\sim$115 MeV) absorption cross sections.

In order to study the nature of many-body structures induced by these compact
two-body structures we study the $dp$, $dd$ and $\alpha d$ overlaps with
the $\Psi_v$ of $^3$He, $^4$He and $^6$Li in section V. These depend
strongly on the spin projection $M_d$ of the deuteron and indicate the
presence of anisotropic structures in all these nuclei. Experiments to
probe these structures are suggested.

Pair distribution functions in other $T,S,M_S$ states are discussed
in section VI. Those in $T,S=1,1$, $M_S=0,\pm1$ states are
anisotropic as expected from the pion-exchange tensor force.
We also find that the number of $T,S=1,0$ pairs in a nucleus is reduced
due to many-body effects involving the strong $T,S=0,1$ tensor correlations.
This reduction gives a significant contribution to the saturation of nuclear
binding energies.

The Skyrme field theory~\cite{Makhankov}, related to QCD in the limit of large
number of colors $N_c\rightarrow\infty$, has predicted toroidal shapes for the
deuteron~\cite{Kopeliovich,Verbaarschot,Leese} in the classical limit.
Density distributions
of the ground states with 3 to 6 baryons have also been
calculated~\cite{Braaten}
in this limit. In the last section, VII, we summarize our results, obtained
with conventional nuclear many-body theory, and indicate their relation to
those of the Skyrme field theory in the classical limit, and of the
constituent quark model.

\section{The Two-Nucleon Interaction in the $T,S = 0,1$ State}

Nuclear forces are not yet quantitatively understood from QCD. However,
many realistic models have been constructed by fitting the available
two-nucleon scattering data. The shape of the short-range structures
in the $T,S = 0,1$ state appears to be relatively model-independent.
The interaction $v_{0,1}$ in the $T,S = 0,1$ state in Reid, Urbana
and Argonne models can be expressed as
\begin{equation}
v_{0,1}=v_{0,1}^c(r)+v_{0,1}^t(r)S_{ij}+v_{0,1}^{ls}(r){\bf L}\cdot{\bf S}
+v_{0,1}^{l2}(r)L^2+v_{0,1}^{ls2}(r)({\bf L}\cdot{\bf S})^2,
\end{equation}
while a $\nabla^2$ operator is used instead of $L^2$ in the Paris potential.
The structures are formed mostly by the static part of the interaction:
\begin{equation}
v_{0,1}^{stat}=v_{0,1}^c(r)+v_{0,1}^t(r)S_{ij}.
\end{equation}
It is instructive to study the expectation values of $v_{0,1}^{stat}$
in eigenstates of the position operator ${\bf r}$ with spin projections
$M_S$=0 and $M_S$=$\pm$1. These depend upon
$r$ and $\theta$, the polar angle of ${\bf r}$ with respect to the
spin-quantization axis $\hat{z}$, and are given by
\begin{eqnarray}
\langle M_S\mbox{=}0|v_{0,1}^{stat}({\bf r})|M_S\mbox{=}0\rangle &=&
v_{0,1}^c(r)-
4v_{0,1}^t(r)\mbox{P}_2(\cos\theta), \\
\langle M_S\mbox{=}\pm 1|v_{0,1}^{stat}({\bf r})|M_S\mbox{=}\pm 1\rangle &=&
v_{0,1}^c(r)+2v_{0,1}^t(r)\mbox{P}_2(\cos\theta).
\end{eqnarray}
The $M_S$=0 expectation value of $v_{0,1}^{stat}$ has the largest variation
with $\theta$ as illustrated in Fig.~\ref{potfig}. The static potential
has a repulsive core; outside the core it is very
attractive for $\theta$=$\pi/2$ and repulsive for $\theta$=0 and $\pi$.
Therefore, in this state the $np$ pairs form a toroidal density
distribution in the $xy$ plane. The potential in the $M_S$=$\pm$1 states is
attractive for $\theta$=0 and $\pi$, and equal to that for $M_S$=0,
$\theta$=$\pi/2$, while it is repulsive for $\theta$=$\pi/2$ and half way
between the $M_S$=0, $\theta$=0 and $\pi/2$ potentials. Thus the
$M_S$=$\pm$1 potential has two distinct minima separated by a barrier,
and therefore the density distributions have a dumbbell shape in this state.

At $r>$1.5 fm, the $v_{0,1}^{stat}$ is dominated by the one-pion-exchange
potential, while at smaller $r$ it has a significant model
dependence. Much of this model dependence is cancelled by the differences
in the momentum-dependent terms in the models.
In particular, the deuteron wave functions calculated from these potentials
have much smaller model dependence.
These are commonly written as
\begin{equation}
\Psi_d^{M_d}({\bf r})=R_0(r)\ {\cal Y}_{011}^{M_d}({\hat r})
                     +R_2(r)\ {\cal Y}_{211}^{M_d}({\hat r}),
\end {equation}
where $R_0(\mbox{=}u/r)$ and $R_2(\mbox{=}w/r)$ are the $S$- and $D$-state
radial wave functions and ${\cal Y}_{LSJ}^M$ are the spin-angle functions.
The $R_0$ and $R_2$ calculated from the different potential models are
shown in Fig.\ \ref{r0r2fig}. The short-range
structures are related to the $R_0$ and $R_2$ functions, and therefore
we expect them to be fairly model-independent.

However, the ``full'' Bonn potential~\cite{Mac87} offers an exception.
The $R_0$ and
$R_2$ predicted by this one-boson-exchange model of the $NN$ interaction
are similar to other predictions at larger $r$, but they have an additional
sharp structure with a range of $\sim$0.2 fm. We will not consider the
possibility of such an additional structure in this work.

\section{The Deuteron}
The short-range structure of the deuteron is most obvious in its density
distribution $\rho_d^{M_d}(r^{\prime},\theta)$ which depends upon the
projection $M_d$ of the total deuteron angular momentum, the distance
$r^{\prime}$ from the deuteron center of mass, and the polar angle $\theta$
of ${\bf r}^{\prime}$; it is independent of the azimuthal angle $\phi$.
Note that the interparticle distance ${\bf r}=2{\bf r}^{\prime}$, and
the standard normalizations,
\begin{eqnarray}
\int_0^\infty r^2\mbox{d}r\left[R_0^2(r)+R_2^2(r)\right] &=& 1, \\
\int\mbox{d}^3{\bf r}^{\prime}\ \rho_d^{M_d}({\bf r}^{\prime}) &=& 2,
\end{eqnarray}
are used in this work.

The $\rho_d^{M_d}({\bf r}^{\prime})$ is given by
$16\Psi_d^{M_d^{\,\dag}}(2{\bf r}^{\prime})\Psi_d^{M_d}(2{\bf r}^{\prime})$,
where the factor 16 comes from the difference in normalizations (3.1) and
(3.2).
A simple calculation using Eq.~(2.5) yields
\begin{eqnarray}
\rho_d^0({\bf r}^{\prime})&=&\frac{4}{\pi}\left[C_0(2r^{\prime})
-2 C_2(2r^{\prime})\mbox{P}_2(\cos\theta)\right] \>\>\>, \\
\rho_d^{\pm 1}({\bf r}^{\prime})&=&\frac{4}{\pi}\left[
C_0(2r^{\prime})+C_2(2r^{\prime})\mbox{P}_2(\cos\theta)
\right] \>\>\>,
\end{eqnarray}
with
\begin{eqnarray}
C_0(r)&=& R_0^2(r)+R_2^2(r) \>\>\> , \\
C_2(r)&=& \sqrt{2} R_0(r)R_2(r)-\frac{1}{2} R_2^2(r) \>\>\>.
\end{eqnarray}

The interesting structure of these density distributions is shown in
Figs.\ \ref{figdeutxz}-\ref{surffig}.
Fig.\ \ref{figdeutxz}  shows $\rho_d^{M_d}(r^{\prime})$ along $\theta=0$ and
$\theta=\pi/2$ directions, noting that
\begin{equation}
\rho_d^0(r^{\prime},\theta\mbox{=}\pi/2) = \rho_d^{\pm
1}(r^{\prime},\theta\mbox{=}0).
\end{equation}
The above densities are the largest while
$\rho_d^0(r^{\prime},\theta\mbox{=}0)$
is the smallest as expected
from the properties of $v_{0,1}^{stat}$ discussed in Sec. II.
The small value of the ratio
$\rho_d^0(r^{\prime},\theta=0)/\rho_d^0(r^{\prime}, \theta=\pi/2)$ indicates
that the deuteron has near maximal tensor correlation at distance
$r^{\prime}<1$ fm or equivalently at $r<2$ fm. This ratio is $\sim$0 for
maximal tensor
correlations.

Figures\ \ref{rhod0} and\ \ref{rhod1} show the density distributions
predicted by the Argonne $v_{18}$ model in the $x^{\prime}z^{\prime}$ plane.
The maximum value of $\rho_d$ is fairly model-independent
(Fig.\ \ref{figdeutxz}) and large ($\sim$0.35 fm$^{-3}$).
The maxima of $\rho_d^0$ (Fig.\ \ref{rhod0}) form a ring with a diameter
of $\sim$1 fm, denoted by $d$, in the $x^{\prime}y^{\prime}$ plane, while
the $\rho_d^{\pm 1}$ has two equal maxima on the $z^{\prime}$-axis
(Fig.\ \ref{rhod1}) at $z^{\prime}$=$\pm d/2$.

The three-dimensional distributions $\rho_d^{M_d}({\bf r}^{\prime})$ can be
obtained by rotating the distributions shown in Figs.\ \ref{rhod0}
and\ \ref{rhod1} about the $z^{\prime}$-axis. They are represented by
equidensity surfaces shown in Fig.\ \ref{surffig} for $\rho_d$=0.24
and 0.08 fm$^{-3}$; all four sections are drawn to the same scale.
The maximum value of $\rho_d^0(\theta\mbox{=}0)$ is
$\sim$0.05 fm$^{-3}$ (Fig.\ \ref{figdeutxz}).  Therefore the equidensity
surfaces for $\rho_d^0$ having $\rho_d>$0.05 fm$^{-3}$ cannot intersect
the $z^{\prime}$-axis, and thus have toroidal shapes shown in
Figs.~\ref{surffig}B and \ref{surffig}D. The central hole in these tori is due
to the repulsive core in $v_{0,1}^{stat}$, and their angular confinement
is due to the tensor force. In absence of the tensor force, $R_2(r) = 0$,
the $\rho_d^0 = \rho_d^{\pm 1}$, and the equidensity surfaces are
concentric spheres.

The maximum value of $\rho_d^{\pm 1}(\theta\mbox{=}\pi/2)$
is $\sim$0.19 fm$^{-3}$ as can be seen from Fig.\ \ref{figdeutxz}. Therefore
the
equidensity surfaces of $\rho_d^{\pm 1}$ for
$\rho_d\geq$0.19 fm$^{-3}$ can not cross the $\theta$=$\pi/2$ plane;
they have two disconnected parts forming a dumbbell as shown in
Fig.\ \ref{surffig}A. At smaller values of $\rho_d$ we also obtain two
equidensity $\rho_d^{\pm 1}$ surfaces (Fig.\ \ref{surffig}C), consisting
of an inner surface due to the repulsive core
enclosed by an outer. At very small $\rho_d$ ($\leq$0.05 fm$^{-3}$) the
equidensity surfaces of $\rho_d^0$ also have disconnected inner
and outer parts, neither close to spherical in shape.

The toroidal shape of the $M_d$=0 equidensity surfaces is more compact
and persists down to smaller $\rho_d$, or equivalently to larger values of
$r^{\prime}$, as can be seen from Figs.\ \ref{figdeutxz}-\ref{surffig}. In the
classical Skyrmion
field theory only this shape is obtained for the distribution of baryon
density in the ground state for two baryons~\cite{Makhankov}.
The deuteron can be considered
to be more deformed in the $M_d$=0 state. For example, the expectation
values of the quadrupole operator $Q({\bf r}^{\prime})=2\> r^{\prime 2}
\mbox{P}_2(\cos\theta)$
obey the relation
\begin{equation}
\int\rho_d^0({\bf r}^{\prime})\ Q({\bf r}^{\prime})\ \mbox{d}^3{\bf r}^{\prime}
=-2\int\rho_d^{\pm 1}({\bf r}^{\prime})\ Q({\bf r}^{\prime})\ \mbox{d}^3
{\bf r}^{\prime}.
\end{equation}

It is rather simple to obtain the $\rho_d^{\pm 1}({\bf r}^{\prime})$
from the $\rho_d^0({\bf r}^{\prime})$. We rotate the
$\rho_d^0({\bf r}^{\prime})$ about the $y^{\prime}$-axis by an angle of
$\pi/2$ so that the toroidal ring is in $y^{\prime}z^{\prime}$ plane
with $x^{\prime}$ as the symmetry axis. This places the deuteron in the
superposition of $M_d$=$\pm$1 states. The $M_d$=$\pm$1 states are obtained by
spinning the rotated toroid about the $z^{\prime}$-axis, and the
$\rho_d^{\pm 1}({\bf r}^{\prime})$ is just the average value of the density
of the spinning toroid, i.e.,
\begin{equation}
  \rho_d^{\pm 1}(r^{\prime},\theta)=\frac{1}{2\pi}\int_0^{2\pi}\ \rho_d^0
  (r^{\prime},\cos^{-1}(\sin\theta \cos\phi))\ \mbox{d}\phi.
\end{equation}
The $L$=1, $M$=$\pm$1 states of the harmonic oscillator, given by
$\phi(r)$=e$^{-\nu r^2}(x\pm$i$y$), are obtained in the same way from the
$\phi(r)$=e$^{-\nu r^2}z$, $L$=1, $M$=0 state.
Therefore it is tempting to consider the toroidal shape of
$\rho_d^0$ as the basic shape of the deuteron. The expectation value
of the current operator is zero in the $M_d$=0 state, therefore one may
regard that as the ``static'' state of the deuteron. Note that the
toroidal shapes cannot be obtained by rotating the dumbbell by $\pi/2$
about the $y^{\prime}$-axis and spinning it about the $z^{\prime}$-axis.
This gives $\rho_d^0(\theta\mbox{=}0,r^{\prime})=
\rho_d^{\pm 1}(\theta\mbox{=}\pi/2,r^{\prime})$ which is not true.
The dumbbell- or cigar-shaped density distribution of the deuteron in the
$M_d$=$\pm$1 state has been studied earlier~\cite{ericson,Blatt}. Unfortunately
the
toroidal distribution of the $M_d$=0 state was not studied, and
its similarity with the predictions of Skyrmion field theory
was not noticed.

The deuteron electromagnetic structure functions $A(q)$ and $B(q)$, and the
tensor polarization $T_{20}(q)$ in elastic electron-deuteron scattering
have been extensively studied
experimentally~\cite{Arn75,Sim81,Cra85,Pla90,Auf85,Arn87,Sch84,The91,Dmi85,Gil90}
and theoretically~\cite{FA76,HT89,Sch91,VNF95}.  They are usually
calculated from the $S$- and $D$-wave
functions $R_0$ and $R_2$ obtained from realistic interactions, by
including in the nuclear electromagnetic current, in addition to the
dominant impulse approximation (IA) operators, relativistic corrections
and two-body meson-exchange contributions~\cite{FA76,Sch91}.  More recently,
calculations
of these observables based on quasipotential reductions of the Bethe-Salpeter
equation and one-boson-exchange interaction models, constrained to fit
nucleon-nucleon data, have also been
carried out~\cite{HT89,VNF95}.  The theoretical
predictions for the structure functions based on both the nonrelativistic
and relativistic approaches are in good agreement with data.
Our interest here is not to improve upon the present theoretical
predictions, but to relate the values of the minima and maxima of
$T_{20}(q)$ and $B(q)$ to the size of the toroidal structure in the
deuteron. The $q$-values of these extrema may be shifted by $\sim$10$\%$
by corrections to the impulse approximation used in the following
pedagogical analysis.

The charge form factors, defined as
\begin{equation}
F_{C,M_d}(q)=\frac{1}{2}\int \rho_d^{M_d}({\bf r}^{\prime})e^{iqz^{\prime}}
\mbox{d}^3{\bf r}^{\prime},
\end{equation}
are shown in Fig.\ \ref{deutlong}. At large $q$ the $F_{C,1}(q)$ gets most
of its contribution
from the two peaks of $\rho_d^1({\bf r}^{\prime})$ (Fig.\ \ref{rhod1}) at
$z^{\prime}$=$\pm d/2$. The Fourier transform of the sum of two
$\delta$-functions
at $z^{\prime}$=$\pm d/2$ is given by $\cos(qd/2)$ with zeros at
$qd$=$\pi$, 3$\pi$, $\cdots$. These zeros are due to the cancellation
of the contribution from the two peaks, and they persist even when the
peaks have a finite width. The first two minima of $F_{C,1}^2(q)$, obtained
from
the Argonne $v_{18}$ $\rho_d^1({\bf r}^{\prime})$, occur at $q_1$=3.6
and $q_2$=12.6 fm$^{-1}$. The effective values $d_i$ estimated from
the minima $q_i$, using $d_i$=$(2i-1)\pi/q_i$, are 0.87 and 0.75 fm for
$i$=1, 2. These values are smaller than the diameter $d$=1 fm because the
dumbbell ends (Fig.\ \ref{surffig}A) are not spherical. Nevertheless
the minima of $F_{C,1}^2$ seem to be primarily
determined by the diameter $d$ of the maximum-density torus.

The Fourier transforms of a disc of thickness $t$, with ${\bf q}$
perpendicular to the disc, are proportional to $\sin(qt/2)/(qt/2)$
irrespective of the shape of the disc. These have zeros at $qt$=2$\pi$,
4$\pi$, $\cdots$, which may be used to obtain the thickness $t$.
The first two zeros of $F_{C,0}(q)$ at $q$=9.2 and 19.5 fm$^{-1}$
(Fig.\ \ref{deutlong}) give values 0.68 and 0.64 fm for
the effective thickness of the torus. The maximum thickness along the $z$-axis
of the calculated equidensity surface at half maximum density is 0.88 fm.

The $T_{20}(q)$ form factor of the deuteron has small magnetic
contributions which depend upon the electron scattering angle $\theta$.
The extrema of $T_{20}$ are not significantly affected by this magnetic
contribution as can be seen from Fig.\ \ref{deutt20}, and ignoring it we obtain
a
rather simple equation:
\begin{equation}
T_{20}(q)\sim
-\sqrt{2}\frac{F_{C,0}^2(q)-F_{C,1}^2(q)}{F_{C,0}^2(q)+2F_{C,1}^2(q)}.
\end{equation}
Its minima occur when $F_{C,1}^2(q)$=0, while the maxima
have $F_{C,0}^2(q)$=0. These minima and maxima correspond to those
values of $q$ at which the recoiling deuteron has only $M_d$=0
or $M_d$=$\pm$1, respectively.  The first minimum of
$T_{20}$ is experimentally known to occur at $q\sim$3.5$\pm$0.5 fm$^{-1}$
in agreement with the value $d\sim$1 fm predicted by realistic potentials.
The first maximum of $T_{20}$ has not yet been experimentally located;
it provides a measure of the thickness $t$.

In magnetic elastic scattering the deuteron spin projection $M_d$ in the
${\bf q}$ direction changes by $\pm$1 since the photon has
$M_{\gamma}$=$\pm$1. Thus the magnetic form factor $F_M(q)$ is a transition
form factor. It has convection current and spin-flip terms~\cite{FA76,Sch91}
of which
the latter is dominant. The $F_M(q)$ calculated with and without
the convection current term are not too different (Fig.\ \ref{deutmq}). The
spin
flip part of $F_M(q)$ can be obtained from the transition density
$\rho_{tr}({\bf r}^{\prime})$:
\begin{eqnarray}
F_M^{sf}(q) &=& (\mu_p+\mu_n)\int e^{iqz^{\prime}}\rho_{tr}({\bf r}^{\prime})
\mbox{d}^3{\bf r}^{\prime}, \\
\rho_{tr}({\bf r}^{\prime}) &=& \frac{2}{\pi}\left\{R_0^2(2r^{\prime})-
\frac{1}{2}R_2^2(2 r^{\prime})-\frac{1}{2}\left[\sqrt{2}
R_0(2 r^{\prime})R_2(2 r^{\prime})+R_2^2(2 r^{\prime})\right]
\mbox{P}_2(\cos\theta)\right\}.
\end{eqnarray}
This transition density is shown in Fig.\ \ref{tranden}; it is dominated
by the toroidal $\rho_d^0({\bf r}^{\prime})$, and its effective thickness
along $z^{\prime}$ axis, obtained from the zeros of $F_M(q)$
(Fig.\ \ref{deutmq}), is $\sim$0.85 fm. The minimum of $F_M(q)$ is
observed~\cite{Arn87} at $q\sim 7$ fm$^{-1}$, supporting the theoretical
estimates of $t$.

The deuteron wave function in momentum space is defined as
\begin{eqnarray}
{\tilde{\Psi}}_d^{M_d}({\bf k}) &=& \frac{1}{(2\pi)^{3/2}}
 \int \mbox{d}^3{\bf r}\ {\rm e}^{-{\rm i}{\bf k}
\cdot {\bf r} }\Psi_d^{M_d}({\bf r}) \nonumber  \\
&=&{\tilde {R}}_0(k)\ {\cal Y}_{011}^{M_d}(\hat{\bf k})
  +{\tilde {R}}_2(k)\ {\cal Y}_{211}^{M_d}(\hat{\bf k}) \>\>\>,
\end{eqnarray}
with
\begin{equation}
{\tilde{R}}_L(k) = {\rm i}^L {\sqrt{2/\pi}} \int_0^\infty dr\ r^2 j_L(kr)
R_L(r)
\>\>\>.
\end{equation}
The momentum distributions ${\tilde {\rho}}_d^{M_d}({\bf k})$, given by
${\tilde {\Psi}}_d^{M_d^{\dag}}({\bf k})
{\tilde {\Psi}}_d^{M_d}({\bf k})$, are then easily obtained as
\begin{eqnarray}
{\tilde {\rho}}_d^0({\bf k})&=&\frac{1}{4 \pi}\left[{\tilde {C}}_0(k)
-2 {\tilde {C}}_2(k)\mbox{P}_2(\cos\theta_k)\right] \>\>\>, \\
{\tilde {\rho}}_d^{\pm 1}({\bf k})&=&\frac{1}{4 \pi}\left[
{\tilde {C}}_0(k)+{\tilde {C}}_2(k)\mbox{P}_2(\cos\theta_k)
\right] \>\>\>,
\end{eqnarray}
where $\theta_k$ is the angle between ${\bf k}$ and $\hat{z}$-axis, and
the ${\tilde {C}}_L(k)$ are defined as in Eqs. (3.5)-(3.6) with
$R_L(r)$ replaced by ${\tilde {R}}_L(k)$.  Note that the $\tilde{\rho}_d$
are normalized such that
\begin{equation}
\int \mbox{d}^3{\bf k}\ {\tilde {\rho}}_d^{M_d} ({\bf k}) = 1 \>\>\>.
\end{equation}

The momentum distributions ${\tilde {\rho}}_d^0(k,\theta_k)$
and ${\tilde {\rho}}_d^{\pm 1}(k,\theta_k)$ for $\theta_k=0$ and $\pi /2$
are shown in Fig.\ \ref{rhodk}.  Note that
\begin{equation}
{\tilde {\rho}}_d^0(k,\theta_k\mbox{=}\pi/2) =
{\tilde {\rho}}_d^{\pm 1}(k,\theta_k\mbox{=}0) \>\>\>.
\end{equation}
The zeros of ${\tilde {\rho}}_d^{\pm 1}(k,\theta_k\mbox{=}0)$
and ${\tilde {\rho}}_d^0(k,\theta_k\mbox{=}0)$ occur at $k \simeq 1.5$
fm$^{-1}$
and $5.2$ fm$^{-1}$, respectively, and are related to the spatial dimensions of
the torus. In naive estimates these minima occur at $\pi/2d$ and $\pi/t$,
respectively. Thus measuring the positions of the zeros in these momentum
distributions would provide an independent estimate of the spatial dimensions
of the toroidal structure in the deuteron.  This information would be
complementary to that yielded by elastic form factors measurements.

The momentum distributions ${\tilde {\rho}}_d^{M_d}(k,\theta_k)$ could in
principle be measured by $(e,e^{\prime}p)$ scattering on polarized
deuterons.  In the one-photon exchange approximation the
$\vec{d}(e,e^{\prime}p)n$ cross section, in the laboratory frame,
is generally expressed as
\begin{eqnarray}
{d^5 \sigma^{M_d} \over dE'_e d\Omega_e' d\Omega_p }&=&
\sigma_{Mott}\ p_pE_p R_{\rm rec} ^{-1}\ \left( v_L R_L +
v_T R_T + v_{LT} R_{LT} + v_{TT} R_{TT} \right) \>\>\>,\\
R_{\rm rec}&=& { E_p E_n \over m^2 } \biggl\vert 1- { E_p p_n \over E_n p_p}
{\hat {\bf p}}_p \cdot {\hat {\bf p}}_n \biggr\vert \>\>\>,
\end{eqnarray}
where $M_d$ is the target spin projection, $E'_e$ and
$\Omega'_e$ are the energy and solid angle of the final
electron, and $\Omega_p$ is the solid angle of the ejected proton.  The
coefficients $v_\alpha$ are defined in terms of the electron
variables, while the structure functions $R_\alpha$ involve
matrix elements
\begin{equation}
\langle n+p;{\bf p}_n {\bf p}_p,M_n M_p \mid O_{{\rm L,T}}({\bf q})
 \mid d,M_d \rangle \>\>\>,
\end{equation}
of the charge ($O_L$) and current ($O_T$) operators between the initial
deuteron and final $n+p$ states. The neutron momentum is
${\bf p}_n$=${\bf q}-{\bf p}_p$, ${\bf q}$ is the momentum transfer,
${\bf p}_m$=--${\bf p}_n$ is the missing momentum, and $M_p$ and $M_n$
are the proton and neutron spin projections, respectively~\cite{Sch91}.
The cross section for unpolarized deuterons,
\begin{equation}
{d^5 \sigma \over dE'_e d\Omega_e' d\Omega_p }=\frac {1} {3} \sum_{M_d=0,\pm 1}
{d^5 \sigma^{M_d} \over dE'_e d\Omega_e' d\Omega_p }\>\>\>,
\end{equation}
has been measured up to $p_m\sim$500 MeV/c, and
there is good agreement between theory and experiment~\cite{Laget}.

In plane-wave-impulse-approximation (PWIA), obtained by neglecting interaction
effects in the final $n+p$ states as well as relativistic corrections and
two-body terms in the charge and current operators, the $M_d$-dependent
cross section is proportional to
\begin{equation}
{d^5 \sigma^{M_d} \over dE'_e d\Omega_e' d\Omega_p }\propto
{\tilde {\rho}}_d^{M_d}({\bf p}_m) \>\>\>.
\end{equation}
Using, for example, tensor polarized deuterium, it should be possible to
measure
the difference between ${\tilde {\rho}}_d^0({\bf p}_m)$ and
${\tilde {\rho}}_d^{\pm 1}({\bf p}_m)$, and therefore
empirically determine the positions of the minima in these momentum
distributions.  Clearly, such an analysis is justified
if the PWIA is valid.  This assumption has been tested by carrying out
the full and PWIA calculations of the $d^5 \sigma^{M_d}
/dE'_e d\Omega_e' d\Omega_p $ in parallel kinematics with $q$ fixed at
500 MeV/c, $\omega$ in the range 290-390 MeV, and the electron
scattering angle $\theta_{\rm e}=10^{\circ}$.  The
results, shown in Fig.\ \ref{2heep}, indicate that, while FSI, two-body
current,
and relativistic corrections are not entirely negligible, at least in the
kinematical region which has been studied here, their effect is small
compared to the difference between the cross sections for $M_d$=0 and $\pm$1.
We therefore conclude that the results of such an experiment could be used
to empirically study the diameter and thickness of the torus.
One might argue that this information could be more easily obtained from
elastic form factors measurements, as discussed above.
However, it should be realized that, in contrast to the
$d(e,e^{\prime})\vec{d}$ data,
the double-coincidence data would allow us to ascertain to what extent
this toroidal structure is due to nucleonic degrees of freedom.

\section{The Two-Nucleon Density Distributions in Nuclei}

The two-nucleon density-distributions in $T$, $S$, $M_T$, $M_S$ two-nucleon
states are defined as
\begin{equation}
\rho_{T,S}^{M_T,M_S}({\bf r}) = \frac{1}{2J+1}\
 \sum_{M_J=-J}^J\langle \Psi^{M_J}|\
 \sum_{i<j} P_{ij}({\bf r},T,S,M_T,M_S)\,|\Psi^{M_J}\rangle \>\>\>.
\end{equation}
Here $|\Psi^{M_J}\rangle$ denotes the ground state of the nucleus with total
angular momentum $J$ and projection $M_J$, and $P_{ij}({\bf r},T,S,M_T,M_S)$
projects out the specific two-nucleon state with ${\bf r}_i-{\bf r}_j={\bf r}$.
For $N = Z$ nuclei, the wave functions used in this study are symmetric
under exchange of neutrons
and protons; hence $\rho_{T,S}^{M_T,M_S}({\bf r})$ is independent of $M_T$.
For $^3$He, we have
\begin{equation}
\rho_{1,S}^{-1,M_S} = 0 ;  \ \ \ \
\rho_{1,S}^{+1,M_S} = 2 \rho_{1,S}^{0,M_S} \>\>\>  ,
\end{equation}
while for larger $N \neq Z$ nuclei, the $M_T$ dependence is nontrivial.
In the following we discuss $\rho_{T,S}^{M_S}$, the average over $M_T$
of $\rho_{T,S}^{M_T,M_S}$.
The $\rho_{T,S}^{M_S}$ is normalized such that
\begin{equation}
\sum_{T,S,M_S}\ (2T+1)\ \int \rho_{T,S}^{M_S}({\bf r})\ \mbox{d}^3{\bf r}\
   =\ \frac{1}{2}A(A-1) \>\>\> ,
\end{equation}
which is the number of pairs in the nucleus. It is a function of $r$ and
$\theta$ independent of the azimuthal angle $\phi$.

It can be verified from Eqs.(3.3-3.6) and (4.1) that in the deuteron
\begin{equation}
\rho_{0,1}^M({\bf r})=\frac{1}{3}\times\frac{1}{16}\ \rho_d^M
({\bf r}^{\prime}={\bf r}/2) \>\>\> .
\end{equation}
Note that the spin-dependent two-body density on the left ($M_S=M$) is an
average
over projections $M_d$ in the deuteron, while the polarized one-body density on
the
right ($M_d=M$) has been summed over spins.
The $\rho_{T,S}^{M_S}$ in $^{3,4}$He, $^{6,7}$Li, and
$^{16}$O have been calculated from variational wave functions using
Monte Carlo techniques.  For the A $\leq$ 7 nuclei, these wave functions
minimize the expectation value of a Hamiltonian consisting of the
Argonne $v_{18}$ two-nucleon and Urbana IX three-nucleon
potentials~\cite{PPCW95}
(for $A$ = 6,7 the minimization is constrained by the experimental rms radii);
a detailed description of the form of the
wave functions is given in Refs.~\cite{APW95,WPP96}. The wave function
for $^{16}$O was obtained
from the variationally best wave function by slightly increasing the
radius of the single-particle part of the wave function so as to reproduce
the experimental rms radius of $^{16}$O.  The details of the $^{16}$O
wave function will be published elsewhere~\cite{Pie96}.
A cluster-expansion including up to four-body clusters with Monte Carlo
integration~\cite{PWP92} was used to compute the two-body densities
in $^{16}$O.

To reduce statistical fluctuations in the calculation, we write
\begin{equation}
\rho_{T,S}^{M_S}({\bf r}) =  \sum_{L=0,2} A_{T,S,L}^{M_S}(r)
    \mbox{P}_L(\cos \theta) \>\>\> ,
\end{equation}
and directly compute the $A(r)$ as
\begin{eqnarray}
A_{T,S,L}^{M_S}(r) &=& \frac{1}{2J+1}\ \frac{2L+1}{4\pi} \times \nonumber \\
 && \sum_{M_J} \int d{\bf R} \Psi^{M_J}({\bf R})^{\dagger}
    \sum_{i<j}\, \frac{1}{r^2_{ij}} \delta(r-r_{ij})\,
     \mbox{P}_L\left({\hat {\bf r}}_{ij}
     \cdot\hat{z}\right) \,P_{ij}(T,S,M_S)\, \Psi^{M_J}({\bf R})  \>\>\> ,
\end{eqnarray}
where ${\bf R}$ represents the coordinates ${\bf r}_1,\ldots,{\bf r}_A$.
Because of the average over the total spin of the nucleus,
the $A_{T,S,L}^{M_S}$ are zero for $L>2$, and
\begin{eqnarray}
A_{T,S=1,L=0}^{M_S=0} & = & A_{T,S=1,L=0}^{M_S=\pm 1} \>\>\> , \\
A_{T,S=1,L=2}^{M_S=0} & = & -2 A_{T,S=1,L=2}^{M_S=\pm 1} \>\>\>  .
\end{eqnarray}
For the deuteron, the $A_{0,1,L}^{M_S}$ are related to the $C_L$ of
Eqs.(3.3-3.6) by $A_{0,1,0}^{M_S} = C_0/48$; $A_{0,1,2}^{1} = C_2/48$.

The shapes of $\rho_{0,1}^{M_S}(r,\theta)$ are very similar at $r\le$ 2 fm
in all the nuclei considered. In order to study the evolution of
$\rho_{0,1}^{M_S}$
with $A$ we divided the $\rho_{0,1}^{M_S}(r,\theta)_A$ by the ratio
$R_{Ad}$ defined as
\begin{equation}
R_{Ad} = \frac{\mbox{Max}\left(\rho_{0,1}^{\pm 1}(r,\theta)_A\right)}
	      {\mbox{Max}\left(\rho_{0,1}^{\pm 1}(r,\theta)_d\right)} \>\>\> .
\end{equation}
The densities so normalized are compared in Fig.\ \ref{rhoA}, and the values of
$R_{Ad}$ are listed in Table I.
Fig.\ \ref{rhoA} shows $\rho_{0,1}^0(r,\theta=0)_A/R_{Ad}$,
$\rho_{0,1}^{\pm 1}(r,\theta=\pi/2)_A/R_{Ad}$, and
$\rho_{0,1}^{0}(r,\theta=\pi/2)_A/R_{Ad}$ for $^2$H, $^4$He, and $^{16}$O.
Note that $\rho_{0,1}^{0}(r,\theta=\pi/2)$ = $\rho_{0,1}^{\pm 1}(r,\theta=0)$
by virtue of equations (4.4), (4.6) and (4.7) in all nuclei.
After normalization by $R_{Ad}$, the various densities for $^3$He lie
between those of $^2$H and $^4$He, while those for $^{6,7}$Li are in between
the $^4$He and $^{16}$O results.  It is obvious from Fig.\ \ref{rhoA} that the
equidensity surfaces of the two-body density $\rho^{M_S}_{0,1}$ are very
similar to those of the deuteron density shown in
figures\ \ref{figdeutxz}-\ref{surffig} at $r<$ 2 fm ($r^{\prime}<$ 1 fm).
At $r<$ 2 fm the ratio $\rho_{0,1}^0(r,\theta=0)/\rho_{0,1}^0(r,\theta=\pi/2)$
is very small, indicating that the tensor correlations have near maximal
strength in all the nuclei considered.  In $^{16}$O the
$\rho_{0,1}^{M_S}$ becomes approximately independent of $M_S$ only for
$r \agt 3$ fm.

Bethe and Levinger suggested in 1950~\cite{Levinger} that at small distances
the relative $T,S$ = 0,1
neutron-proton wave function in a nucleus is likely to be similar
to that in the deuteron. We find that this is a good approximation.
The expectation value of any short-ranged two-body
operator that is large only in the $T,S = 0,1$ state scales as
$R_{Ad}$. In Table I we list values of the ratios of the
calculated expectation values of the one-pion exchange part of
the Argonne $v_{18}$ potential, the observed low-energy (118 MeV
for $^3$He~\cite{3Hepion} and $^4$He~\cite{4Hepion},
and 115 MeV for $^{16}$O~\cite{16Opion}) pion
absorption cross sections and the average value of the observed photon
absorption cross sections in the range $E_{\gamma}$ = 80 to 120 MeV.
All these processes are dominated by the $T,S = 0,1$ pairs, and seem
to scale as $R_{Ad}$.

While comparing these ratios in detail it should be realized that
$\langle v_{\pi}\rangle$ in nuclei has a relatively
small contribution from $T,S\neq 0,1$ states, absent in the deuteron, which
makes $\langle v_{\pi}\rangle_A/\langle v_{\pi}\rangle_d$ slightly larger
than $R_{Ad}$. The $\langle v_{\pi}\rangle_d$ = --21.3 MeV for the
Argonne $v_{18}$ model, and it accounts for most of the deuteron potential
energy, $\langle v\rangle_d$ = --22.1 MeV. Also in larger nuclei, the
$\langle v_{\pi}\rangle_A$ gives a large fraction of the total two-body
interaction energy~\cite{PWP92}. Direct comparison of the ratio of pion
absorption cross sections with $R_{Ad}$ may not be strictly valid.
The scattering and absorption of pions by spectator nucleons, absent in the
deuteron, is expected to reduce the ratio $\sigma_{ab,A}^{\pi}/
\sigma_{ab,d}^{\pi}$, while three-body and higher absorption mechanisms,
also absent in the deuteron, will increase it. After correcting for final
state interactions of the two outgoing protons, the two-body
($\pi^+,p\,p$) part is estimated to account for $\sim 76\%$ of the
total absorption cross section for 115 MeV $\pi^+$ by $^{16}$O~\cite{16Opion}.
In $^3$He $\sim 20\%$ of the 118 MeV $\pi^+$ absorption cross section
has three-body character~\cite{3Hepion}, however a part of this $20\%$ must
be due to initial and final state interactions.

Results of Mainz experiments~\cite{Mainz} on $^7$Li and $^{16}$O are used to
calculate the average value of $\sigma^{\gamma}$ in the energy interval
$E_{\gamma}$ = 80 to 120 MeV. The $\sigma_{ab,d}^{\gamma}$ averaged over
the same energy interval is $\sim 0.072$ mb~\cite{photo-deuteron}.
The only available data for $^3$He in this energy range are from the
experiments done in
the 1960's~\cite{3he3bodyphoto} and 1970's~\cite{3he2bodyphoto}. The
average cross section of the two-body photodisintegration of $^3$He,
in the energy range 80-120 MeV, is $\sim$0.03 mb~\cite{3he2bodyphoto}, and
that for the three-body process is $\sim$0.10 mb~\cite{3he3bodyphoto},
giving total cross section of $\sim$0.13 mb. The average cross section for
total absorption of photons by $^4$He in the same energy range is crudely
estimated from Fig.\ \ref{potfig} in ref.~\cite{4hephoto} to be $\sim$0.3 mb.

The total number of pairs with given $T,S$ in nuclei can be computed as
\begin{eqnarray}
N^{A}_{T,S} &=& \sum_{M_S}\ (2T+1)\  2 \pi \int r^2dr \  d\cos \theta \
  \rho^{M_S}_{T,S}(r,\theta)_A \nonumber  \\
&=& (2T+1)(2S+1)\ 4 \pi \int r^2dr \  A^{0}_{T,S,0}(r) \>\>\>;
\end{eqnarray}
the values for $T,S = 0,1$ and the corresponding naive independent-particle
model values are also shown in Table I.
We see that the correlations induced by the potentials do not significantly
change the $N^{A}_{0,1}$ from their independent-particle (IP) values; however,
as will be discussed later, this is not true for $T$=1 pairs.
For few-body nuclei, $R_{Ad}$ is significantly larger than $N^{A}_{0,1}$,
however, in $^{16}$O  $N^{A}_{0,1}$ has a large contribution from pairs
with large $r$ and  $R_{Ad}$ is smaller than  $N^{A}_{0,1}$.  The calculated
value of $R_{Ad}$ for $^{16}$O is much smaller than Levinger's estimate
$R_{Ad}\sim 8NZ/A$~\cite{Levinger}.

The $\rho_d^{M=0}(r,\theta=\pi/2)$ has its half-maximum value at $r\sim$1.8 fm
(Fig.\ \ref{rhoA}). If we identify the region with $r<1.8$ fm as
the ``quasi-deuteron'',
then the probability that the $np$ pair in a deuteron is in the
quasi-deuteron region is $\sim$0.25, and the number of quasi-deuterons in
a nucleus is $\sim R_{Ad}/4$. In the past, however, $R_{Ad}$ itself has been
interpreted as the number of quasi-deuterons in the nucleus.

\section{Two-Cluster Distribution Functions}

The strong spin-dependent anisotropies of the two-nucleon densities
suggest that three-nucleon and higher distribution functions in nuclei
could also be anisotropic.
A general study of these higher distributions is beyond the scope of this work;
however the two-cluster distribution functions $\vec{d} \vec{p}$ in $^3$He,
$\vec{d} \vec{d}$ in $^4$He, and $\alpha \vec{d}$ in $^6$Li are simple
to study with the Monte Carlo method~\cite{MC}.
They provide some information on the higher distribution functions, and may
be relatively accessible by $(e,e^{\prime}\vec{d})$ and $(e,e^{\prime}\vec{p})$
experiments.

The two-cluster overlap function can be written generally as
\begin{eqnarray}
A_{ab}(M_a,M_b,M_J,{\bf r}_{ab}) &=& \langle {\cal A} \Psi^{M_a}_a \Psi^{M_b}_b
  , {\bf r}_{ab} | \Psi^{M_J} \rangle  \\
  &=& \sum_{L M_L S M_S} \langle L M_L S M_S | J M_J \rangle
   \langle J_a M_a J_b M_b | S M_S \rangle R_L(r_{ab}) Y_{LM_L}(\hat{r}_{ab})
   \nonumber \ ,
\end{eqnarray}
where ${\bf r}_{ab}$ is the is the relative coordinate between the centers of
mass of the two clusters and ${\cal A}$ is an antisymmetrization operator
for the two-cluster state.
The $R_L(r_{ab})$ radial functions can be evaluated from
\begin{eqnarray}
R_L(r_{ab}) &=& \sum_{M_a M_b M_L M_S} \langle J_a M_a J_b M_b | S M_S \rangle
           \langle LM_L SM_S|JM_J \rangle \nonumber \\
       && \int d{\bf R} [{\cal A} \Psi^{M_a}_a({\bf R}_a)
          \Psi^{M_b}_b({\bf R}_b)]^{\dagger} Y^*_{LM_L}(\hat{r}_{ab})
          \frac{\delta(r-r_{ab})}{r^2_{ab}} \Psi^{M_J}({\bf R}) \ ,
\end{eqnarray}
where ${\bf R}_{a(b)}$ represents the coordinates of particles in cluster
$a(b)$. We note that in PWIA the $(e,e^{\prime}\vec{a})\vec{b}$ cross
section is proportional to the momentum distribution
$|\tilde{A}_{ab}(M_a,M_b,M_J,{\bf k})|^2$ obtained from the Fourier
transform of the overlap function $A_{ab}(M_a,M_b,M_J,{\bf r}_{ab})$.

In the present work, the integrations have been made with Monte Carlo
techniques akin to Ref.\cite{MC}, but with some improvements.
Configurations are sampled with the weight function $|\Psi_v^{M_J}|^2$
containing the full variational wave function. In Ref.\cite{MC} only a
single term in the antisymmetric product in Eq.~(5.2) is calculated.
The efficiency of the Monte Carlo sampling has been improved by evaluating
 all possible partitions
of the nucleus into clusters $a$ and $b$ at each configuration ${\bf R}$.
We also use a much larger sample of configurations than in the previous
calculations.

We can define a two-cluster wave function,
in analogy with the deuteron wave function of Eq.~(2.5), using the radial
overlap functions
\begin{eqnarray}
\Psi^{M_J}_{ab}({\bf r}_{ab}) &=& \sum_{LS} R_L(r_{ab})
              {\cal Y}^{M_J}_{LSJ}(\hat{r}_{ab}) \nonumber \\
  &=& \sum_{M_a M_b} A_{ab}(M_a,M_b,M_J,{\bf r}_{ab})
     \chi^{M_a} \chi^{M_b}  \ ,
\end{eqnarray}
where $\chi^{M_a}$ and $\chi^{M_b}$ denote spin states $J_aM_a$ and
$J_bM_b$ of $a$ and $b$.
For the cases $ab$ = $dp$, $dd$, and $\alpha d$ there are both S-
and D-wave states in the two-cluster wave function.
In these cases the well known $D_2$ parameter can be defined by means of the
$R_0(r)$ and $R_2(r)$ radial functions~\cite{ES90}:
\begin{equation}
   D^{ab}_2 = \frac{\int R_2(r_{ab}) r_{ab}^4 dr_{ab}}
           {15 \int R_0(r_{ab}) r_{ab}^2 dr_{ab}} \ .
\end{equation}
Although in the present paper we emphasize the short-range
structure of nuclei, it is also interesting to study the asymptotic
behavior of the overlap radial functions.
Of particular interest is the asymptotic D/S ratio
$\eta_{ab} = C^{ab}_2 / C^{ab}_0$,
where $C_0$ and $C_2$ are the asymptotic normalization constants of $R_0(r)$
and $R_2(r)$ respectively:
\begin{equation}
   R_L(r_{ab}) = \lim_{r_{ab} \rightarrow \infty}
                 -{\rm i}^L C^{ab}_L h_L( {\rm i} \alpha_{ab} r_{ab} ).
\end{equation}
Here $h_L$ is the spherical Hankel function of first kind and $\alpha_{ab}$ is
the wave number associated with the separation energy of the nucleus into
clusters $a$ and $b$.
We must point out that the present variational method, as well as the GFMC
method, determine the wave functions by energy minimization, to which
long-range configurations contribute very little. Therefore these methods are
not very sensitive to the asymptotic part of the wave functions, and
consequently our values for $\eta_{ab}$ should be considered only as estimates.

The two-cluster density distribution for a given set of spin projections
is defined as:
\begin{equation}
   \rho_{ab}^{M_a,M_b,M_J}({\bf r}_{ab}) = |A_{ab}(M_a,M_b,M_J,{\bf
r}_{ab})|^2.
\end{equation}
In each of the cases studied here, it exhibits spin-dependent spatial
anisotropies which are easily understood in terms of
the toroidal or dumbbell structure of the polarized deuteron.  The density
is enhanced in the direction corresponding to the most efficient or compact
placement of the deuteron with the remaining cluster, and reduced in those
directions that would lead to very extended structures.

Finally, we are also interested in the total normalizations $N_L$ of the S- and
D-wave two-cluster distributions:
\begin{equation}
   N^{ab}_L = \int_0^\infty r_{ab}^2 dr_{ab} \ R_L^2(r_{ab}) \ .
\end{equation}
These quantities can be related to spectroscopic factors and give the total
S- and D-state fractions.
All the results presented here are obtained from the Argonne $v_{18}$
two-nucleon and Urbana model IX three-nucleon interactions.

\subsection{$\vec{d} \vec{p}$ Distribution in $^3$He}

The calculated $R_0(r_{dp})$ and $R_2(r_{dp})$ are shown in
Fig.\ \ref{R02dp}; the $R_2(r_{dp})$ is negative and smaller in magnitude
than the $R_2$ in deuteron (Fig.\ \ref{r0r2fig}).
The $D^{dp}_2$ value obtained with these radial functions is
$-0.15 \pm 0.01$ fm$^2$, a little smaller than experimental estimates,
ranging from $-0.20 \pm 0.04$ to $-0.25 \pm 0.04$, obtained through
DWBA analysis of ($d$,$^3$He) transfer reactions~\cite{ES90}.
In Fig.\ \ref{R02dp} we also show our asymptotic fit to the S- and D-waves.
The result is $\eta_{dp} = -0.035$, somewhat smaller than the Faddeev
result~\cite{FGLP88}, $-0.043 \pm 0.001$.
Experimental estimates, also obtained through DWBA analysis of ($d$,$^3$He)
transfer reactions, range from $-0.042 \pm 0.007$ to
$-0.035 \pm 0.006$~\cite{ES90}.

The total normalizations are $N^{dp}_0 = 1.31$ and $N^{dp}_2 = 0.022$.
The sum, 1.33, can be interpreted as the number of deuterons in
$^3$He~\cite{MC}.
It is less than 1.49 (Table I), the number of $T,S=0,1$ pairs, because the
pairs are not always in the deuteron state.
It is also smaller than $R_{Ad} = 2.0$ inferred from short range
distribution functions (Fig.\ \ref{rhoA} and Table I).
This is probably because $^3$He is more compact than the deuteron.

The $\rho_{dp}^{0,\case{1}{2},\case{1}{2}}$ and
$\rho_{dp}^{1,-\case{1}{2},\case{1}{2}}$ are shown
in Figs.\ \ref{3Hemd0} and\ \ref{3Hemd1}.
When $M_d=0$, the $M_p=+\case{1}{2}$ proton is
preferentially along the $z$-axis; in contrast, when
$M_d=1$, the $M_p=-\case{1}{2}$ proton is more likely to be in the $xy$ plane.
In the first density distribution the S- and D- wave amplitudes interfere
constructively, to enhance the probability of finding the proton along the
$z$-axis, whereas in the latter the interference is constructive on
the transverse $xy$-plane.
Consequently $R_0$ and $R_2$ have opposite signs and $D^{dp}_2$ and
$\eta_{dp}$ are both negative. The spin-dependent $\vec{d}\vec{p}$
anisotropies are
favored by both tensor and central forces, and lead to more compact
three-body states.

The momentum distribution of $dp$ clusters in $M_J=\case{1}{2}$ $^3$He is
shown in Fig.\ \ref{3Hemdk} for $M_d$, $M_p$ = 0, $\case{1}{2}$ and 1,
$-\case{1}{2}$ for momenta parallel and transverse to the $z$-axis. In PWIA the
$^3\vec{\mbox{He}}(e,e^{\prime}\vec{\mbox{p}})d$ cross-section is
directly related to these momentum distributions. A large spin dependence of
the missing-momentum distribution for protons
ejected in parallel kinematics is predicted for ${\bf q}$ parallel
to $\hat{z}$. The minimum for the momentum distribution along $z$-axis
for $M_p=+\case{1}{2}$ occurs at $\sim$1.4 fm$^{-1}$, while that for
$M_p=-\case{1}{2}$
is at $\sim$2.4 fm$^{-1}$. Thus the spin asymmetry,
$(n_{\uparrow}-n_{\downarrow})/(n_{\uparrow}+n_{\downarrow})$,
of the protons ejected from
polarized $^3$He changes from $\sim$ --1 to +1 as the missing momentum
varies from $\sim$1.4 to 2.4 fm$^{-1}$ in parallel kinematics and PWIA.
The $dp$ momentum distribution in unpolarized $^3$He has been studied
at Saclay~\cite{Marchand} up to $\sim$2.5 fm$^{-1}$. The observed distribution
is generally smaller than the PWIA prediction~\cite{Laget} indicating
attenuation due to FSI. However, a part of the FSI attenuation will cancel in
the asymmetry, and moreover, it is now possible to perform continuum
Faddeev calculations~\cite{Glockle} including FSI.

\subsection{$\vec{d} \vec{d}$ Distribution in $^4$He}

The calculated $R_0(r_{dd})$ and $R_2(r_{dd})$ are shown in Fig.\ \ref{R02dd}.
The $D^{dd}_2$ value obtained with these radial functions is
$-0.12\pm0.01$ fm$^2$.
In Fig.\ \ref{R02dd} we also show our asymptotic fit to the S- and D-waves.
The result is $\eta_{dd} = -0.091$.
The integrals of these functions yield $N^{dd}_0 = 0.98$ and
$N^{dd}_2 = 0.024$.
The number of deuterons present is greater than twice the sum of these
quantities, 2.0, when one allows for the additional presence of $d+p+n$
configurations.
The $\rho_{dd}^{0,0,0}$ and $\rho_{dd}^{1,-1,0}$ are large in $^4$He
and their anisotropies, induced by the tensor interaction and the shapes
of deuterons,
are shown in Fig.\ \ref{4Hedd}.
The $\rho_{dd}^{0,0,0}$ is largest when $r_{dd}$ is along the $z$-axis, i.e.,
when the deuterons are in the toroidal shape and have a common axis. It is
smallest when ${\bf r}_{dd}$ is transverse (two tori side by side) and equal
to that for $\rho_{dd}^{1,-1,0}$ with ${\bf r}_{dd}$
parallel to $\hat{z}$ (two dumbbells in a line). The latter distribution
is of intermediate strength when ${\bf r}_{dd}$ is transverse (two dumbbells
side by side).
Again in the first (second) density distribution the S- and D-wave amplitudes
interfere constructively (destructively) along the $\hat{z}$-axis.
Therefore $R_0$ and $R_2$
have opposite signs and $D^{dd}_2$ and $\eta_{dd}$ are both negative.

The momentum distributions are also anisotropic
(Fig.\ \ref{4Heddk}).
In particular the $\tilde{\rho}_{dd}^{0,0,0}(k\hat{z})$ has a dip at
$k\sim$1.7 fm$^{-1}$ that is absent in the
$\tilde{\rho}_{dd}^{1,-1,0}(k\hat{z})$.
It may be possible to study these with
$(e,e^{\prime}\vec{d})$ reactions. The unpolarized $^4$He$(e,e^{\prime}d)d$
reaction has been studied at NIKHEF~\cite{Ent}. The observed cross sections
are much smaller than estimates using the $dd$ momentum distribution and either
PWIA or DWIA.

\subsection{$\alpha \vec{d}$ Distribution in $^6$Li}

The calculated $R_0(r_{\alpha d})$ and $R_2(r_{\alpha d})$ are shown in
Fig.\ \ref{R02ad}.
The $R_0(r_{\alpha d})$ and $R_2(r_{\alpha d})$ both exhibit nodes at
short distances and have opposite signs almost everywhere.
This nodal structure has been predicted in $\alpha + d$ and $\alpha + p + n$
cluster models, but not always with the correct relative sign~\cite{KPRVR95}.
The asymptotic behavior is correlated with the quadrupole moment $Q$;
obtaining the experimental value of $-0.08$ fm$^2$ has been a long-standing
problem in $\alpha + p + n$ cluster models.
The variational wave function used here gives $Q = -0.8\pm 0.2$ fm$^2$, i.e.,
the correct sign but far too large in magnitude.
Small changes in the long range part of the $^6$Li wave function have
effects of order 1 fm$^2$ on the quadrupole moment. Thus the values of
the asymptotic properties, $D_2^{\alpha d}$=--0.29 fm$^2$ and
$\eta_{\alpha d}$=--0.07$\pm 0.02$, obtained with this wave function
may not be very accurate.  The total normalizations are
$N^{\alpha d}_0 = 0.82$ and $N^{\alpha d}_2 = 0.021$.
The resulting spectroscopic factor, 0.84, is in good agreement with the value
of 0.85 obtained in radiative capture experiments~\cite{Robertson81}.

The two-cluster densities $\rho_{\alpha d}^{0,M_d,M_J}({\bf r})$, multiplied by
$r^2$, are shown in Fig.\ \ref{6Liad}. They have two peaks; the smaller inner
peak at $r\sim 0.9$ fm is almost spherically symmetric, while the larger peak
at $r\sim 4$ fm is anisotropic.
In particular $\rho_{\alpha d}^{0,0,0}(r\hat{z})$ is much larger than
$\rho_{\alpha d}^{0,0,0}(r\hat{x})$ for $r > 2$ fm.
In the former configuration the $r$ is along the axis of the torus,
while in the latter it is transverse. This anisotropy is also a consequence
of the toroidal shape of the deuteron in the $M_d$=0 state.

\section{Other $T,S$ Channels}

In this section we discuss the properties in nuclei of pairs of nucleons
with $T,S = 0,0$, $1,0$, and $1,1$.
Like the $T,S = 0,1$ channel discussed in the previous sections,
the $T,S = 1,1$ channel also has a tensor potential, but it has the
opposite sign of that for $T,S = 0,1$.  Therefore
the role of $M_S$ is reversed compared to that in $T=0$ states; $M_S = 0$
pairs have maximum density along the $z$-axis, while $M_S = \pm 1$ pairs
have maximum density in the $xy$ plane as can be seen in
Fig.\ \ref{ts11fig}, which shows
$\rho^{M_S}_{1,1}(r,\theta)/R^A_{1,1}$ for $^4$He, $^6$Li, and $^{16}$O.
The curves for $M_S =\pm 1$, $\theta = \pi/2$ are between the two sets of
curves shown in the figure;
to reduce clutter they are not shown.  The curves for $^4$He and $^6$Li
have been renormalized by the factors $R^A_{1,1}$ to have the same
peak height as for $^{16}$O; these factors are shown in Table II.
We see that the shapes of the $T,S = 1,1$ density profiles are quite
different in the different nuclei and that the anisotropy decreases
as the number of nucleons increases.

The analog of the deuteron in the $T,S = 1,0$ channel is the
$^1$S$_0$ virtual bound state (VBS).  For the Argonne $v_{18}$ potential,
this is a pole on the second energy sheet at $E = -0.098$ MeV or
$k = -0.049 {\rm i}$ fm$^{-1}$. Although the wave function is not
normalizable, it has a local peak which we scale to compare to the
unpolarized deuteron in Fig.\ \ref{ts10fig}.
We see that it peaks at a slightly larger radius
and is broader.  The figure also shows the
$\rho^0_{1,0}(r)/R^A_{1,0}$ of $^4$He, $^6$Li, and $^{16}$O;
the curve for $^3$He is between those of $^4$He and $^6$Li,
while the curve for $^7$Li is very close to that of $^6$Li.
Again the curves have been normalized to the peak height of the $^{16}$O
density.
We see that the short-range
shapes of the $\rho^0_{1,0}$ in nuclei are well reproduced by the
VBS density.
Finally, Fig.\ \ref{ts00fig} shows the densities for the $T,S = 0,0$
channel, again normalized to $^{16}$O. As is the case for $T,S = 1,1$,
there is no common shape.

Table II also shows the number of pairs, $N^A_{T,S}$,
in these $T,S$ channels
and the corresponding IP values.  As is the case for $T,S = 0,1$
(Table I), the number of pairs increases more rapidly with A than does
$R^A_{T,S}$, because of the increasing proportion of pairs with
large separation.

Using the projection operators
$(1-\mbox{\boldmath$\tau$}_i\cdot\mbox{\boldmath$\tau$}_j)/4$ and
$(3+\mbox{\boldmath$\tau$}_i\cdot\mbox{\boldmath$\tau$}_j)/4$ for
$T=0$ and 1 pairs we find that the total
number of $T=0$ and 1 pairs in a nucleus depends only on its mass
number $A$ and isospin $T_A$:
\begin{eqnarray}
N_{0,0}^A+N_{0,1}^A &=& \frac{1}{8}\left[\ \,A^2+2A-4T_A(T_A+1)\right], \\
N_{1,0}^A+N_{1,1}^A &=& \frac{1}{8}\left[3A^2-6A+4T_A(T_A+1)\right].
\end{eqnarray}
The above relations are obeyed by $N_{T,S}^A$ obtained from either the
$IP$ or correlated wave functions, since, in the present study,
both are eigenstates of $T_A$.

If the total spin,
\begin{equation}
{\bf S}_A=\sum_i \frac{1}{2}\mbox{\boldmath$\sigma$}_i \, ,
\end {equation}
were to be a good quantum number we would have similar relations,
\begin{eqnarray}
N_{0,0}^A+N_{1,0}^A &=& \frac{1}{8}\left[\ \,A^2+2A-4S_A(S_A+1)\right], \\
N_{0,1}^A+N_{1,1}^A &=& \frac{1}{8}\left[3A^2-6A+4S_A(S_A+1)\right],
\end{eqnarray}
for the total number of pairs with spin 0 and 1. They are obeyed by the
$N_{T,S}^A$ calculated for the $IP$ states which have $S_A = 1, \case{1}{2},
0, 1, \case{1}{2}$
and 0 for $^2$H, $^3$He, $^4$He, $^6$Li, $^7$Li and $^{16}$O respectively.
However, $S_A$ is not a good quantum number; tensor correlations admix
states with larger $S_A$ in the ground state. These reduce the $N_{1,0}^A$
and increase the $N_{1,1}^A$ by the same amount due to Eq.~(6.2). In
$^3$He ($^4$He) the $N_{1,1}^A$ is given by 1.5$P_D$ (3$P_D$), where
$P_D$ is the fraction of $L=2$, $S_A=\case{3}{2}$ ($S_A=2$) state in the
nuclear
ground state.

The interaction in the $T,S = 1,0$ state is much more attractive
than that in the $T,S = 1,1$ state. Hence the depletion of $T,S = 1,0$
pairs by tensor correlations reduces the binding energy of nuclei
significantly. For example, in $^4$He the $T,S = 1,0$ interaction gives
$-14.2$ MeV per pair, while the $T,S = 1,1$ interaction gives only --0.8 MeV
per pair. Thus the conversion of 0.47 $T$=1 pairs from $S$=0 to $S$=1
state raises the energy of $^4$He by $\sim 6.3$ MeV.
It should be stressed that this is a ``many-body'' effect absent
in the two-body cluster approximation of either Brueckner or variational
methods. The tensor interaction between nucleons $i$ and $j$ can flip their
spins and convert pairs $ik$ and/or $jl$ from $S$=0 to $S$=1.

\section{Conclusions}

The main conclusions of this study of nuclear structure, as predicted by
realistic models of nuclear forces, are:

(i) The static part of the two-nucleon potential in $T,S,M_S = 0,1,0$
state has a large angular dependence due to the tensor interaction dominated
by one-pion exchange. At $r\sim 1$~fm the difference between this potential
at $\theta$ = $\pi/2$ and 0 is $\sim 300$ MeV in most models
(Fig.\ \ref{potfig}). It
confines $T,S,M_S = 0,1,0$ pairs to the small $\theta$ region producing
toroidal distributions. The central hole in these tori is due to the repulsive
core in $NN$ interaction. The maximum density in the tori is large, due to
which the peak one-body density in deuterium exceeds 0.3 fm$^{-3}$ in most
models.

(ii) The more familiar dumbbell (or cigar) shaped density distribution of the
deuteron in $M_S$=$\pm 1$ states can be considered as that produced by a
rotating torus.

(iii) The diameter of the maximum density torus, and the thickness of the
half-maximum density torus are predicted to be $\sim 1.0$ and 0.9 fm,
respectively; these values are supported by the observed elastic
electron-deuteron scattering.

(iv) The pair distribution functions in $T,S = 0,1$ states indicate that
the tensor correlations have near maximal strength in all nuclei considered
here for $r\le 2$ fm.

(v) The pair distribution functions in $T,S = 0,1$ and 1,0 states in
different nuclei, can be scaled to lie on universal surfaces for $r\le 2$ fm.
These universal surfaces are predicted by the density distributions of the
deuteron and the $^1S_0$ virtual bound state. The scaling factor $R_{Ad}$
for the $T,S = 0,1$ densities provides a rigorous definition of the
Levinger-Bethe quasi-deuteron number of the nucleus.  The calculated values
of $R_{Ad}$ are significantly different from estimates based on
independent-particle models, and in qualitative agreement with photon and pion
absorption data.

(vi) The many-body distribution functions are also predicted to be anisotropic.
In particular the anisotropies of the $\vec{d}\vec{p}$, $\vec{d}\vec{d}$, and
$\alpha\vec{d}$ distributions in $^3$He, $^4$He and $^6$Li are strongly
influenced
by the toroidal structure of the deuteron.

(vii) Tensor correlations convert $T$=1 pairs of nucleons from $S$=0 to $S$=1
state. This many-body effect reduces the binding energies of nuclei. It
does not appear as if many-body effects reduce the magnitude of tensor
correlations for the range of nuclei studied here: $^2$H to $^{16}$O.

Due to the small size of this toroidal structure it may be worthwhile to
attempt
to understand it from the more basic quark degrees of freedom. Within the
constituent quark model~\cite{Isgur,Carlson} it requires a solution of
the six-quark
problem with a suitably chosen Hamiltonian. Many attempts have been made (see
Refs.~\cite{Robson,FVSF,Yu} for example) to calculate the
nucleon-nucleon interaction from approximate solutions of the six-quark
Hamiltonian. A direct coupling of the pions to the quarks is used to obtain
the tensor part of the interaction. The toroidal structure
is presumably very sensitive to this coupling and to the tensor part of the
quark-quark interaction in the framework of the constituent quark model.

As is well known, toroidal structure for the ground state of the deuteron
was predicted many years ago~\cite{Kopeliovich,Verbaarschot} using
classical Skyrme field theory~\cite{Makhankov} related to QCD in the
$N_c\rightarrow\infty$ limit.
In the classical limit one obtains a toroidal shape of $\sim 1$ fm in size
and a binding energy of $\sim 150$ MeV. From Fig.\ \ref{potfig} it
is obvious that
in the classical limit realistic models of nuclear forces would also give
a deuteron binding energy in the 100 to 200 MeV range. There have been
attempts to include quantum corrections to this theory. A recent
calculation~\cite{Leese} obtains an energy of --6.18 MeV for the deuteron
in this model. Ground states of the classical Skyrme field with baryon
numbers 3 to 6 have also been studied~\cite{Braaten}. The baryon
equidensity surfaces of these classical solutions are highly
anisotropic. However, the nucleon equidensity surfaces of the
$J^{\pi}=\case{1}{2}^+$ and $0^+$ $^3$He and $^4$He must be spherically
symmetric, thus a direct comparison is not possible. Nevertheless
the anisotropic $\vec{d}\vec{p}$ and $\vec{d}\vec{d}$ distributions
in $^3$He and $^4$He may have some relation to the baryon density
distributions in the Skyrme model.

\acknowledgements

The authors wish to thank Brian Pudliner for many interesting discussions.
A.A. wishes to thank A. M. Eir\'{o} for useful
discussions on the two-cluster overlap functions. R.B.W. wishes to thank
Dieter Kurath for useful comments on constructing wave functions for
$^{6,7}$Li.
The calculations were made possible by grants of computer time from the
Mathematics and Computer Science Division of Argonne National Laboratory,
the Pittsburgh Supercomputing Center, the Cornell Theory Center and the
National Energy Research Supercomputer Center.
The work of J.L.F, V.R.P. and A.A. has been
partially supported by U.S. National Science Foundation via Grant No.
Phy94-21309, that of S.C.P. and R.B.W. by the U.S. Department of Energy,
Nuclear Physics Division under contract No. E-31-109-ENG-38, that of A.A.
by Universidade de Lisboa, Junta de Investiga\c c\~ao Cient\'{\i}fica e
Tecnol\'{o}gica under contract No. PBIC/C/CEN/1108/92, and that of R.S.
by the U.S. Department of Energy.

\newpage

\begin{figure}
\caption{The upper four lines show expectation values of $v_{0,1}^{stat.}$
for $M_S$=0, $\theta$=0, and the lower four lines are for $M_S$=0,
$\theta$=$\pi$/2 or equivalently $M_S$=$\pm$1, $\theta$=0. The expectation
values for $M_S$=$\pm$1, $\theta$=$\pi$/2 (not shown) are half way in between.}
\label{potfig}
\end{figure}

\begin{figure}
\caption{The $S$- and $D$-wave deuteron wave functions for various potential
models.}
\label{r0r2fig}
\end{figure}

\begin{figure}
\caption{The top, middle and bottom four curves respectively show the
deuteron density for the indicated values of $M_d$ and $\theta$, obtained from
various potential models.}
\label{figdeutxz}
\end{figure}

\begin{figure}
\caption{The deuteron density $\rho_d^0(x^{\prime},z^{\prime})$ obtained
from the Argonne $v_{18}$ model. The peaks are located at $z^{\prime}$=0
and $x^{\prime}$=$\pm$d/2.}
\label{rhod0}
\end{figure}

\begin{figure}
\caption{The deuteron density $\rho_d^{\pm 1}(x^{\prime},z^{\prime})$ obtained
from the Argonne $v_{18}$ model. The peaks are located at $x^{\prime}$=0
and $z^{\prime}$=$\pm$d/2.}
\label{rhod1}
\end{figure}

\begin{figure}
\caption{The surfaces having $\rho_d^{\pm 1}({\bf r}^{\prime})$=0.24 fm$^{-3}$
(A) and $\rho_d^0({\bf r}^{\prime})$=0.24 fm$^{-3}$ (B). The surfaces
are symmetric about $z^{\prime}$ axis and have $r^{\prime}\le$0.74 fm, i.e.,
the length of the dumbbell along $z^{\prime}$ axis as well as the diameter of
the outer surface of the torus is 1.48 fm. Sections C and D are for
$\rho_d^{M_d}({\bf r}^{\prime})$=0.08 fm$^{-3}$; the maximum value of
$r^{\prime}$ is 1.2 fm.}
\label{surffig}
\end{figure}

\begin{figure}
\caption{The square of the calculated deuteron charge form factors.}
\label{deutlong}
\end{figure}

\begin{figure}
\caption{The values of deuteron $T_{20}(q)$ obtained from Eq.~(3.11) are
shown by full line, whereas the dashed line gives $T_{20}(q)$ including
magnetic contributions for a 15$^{\circ}$ electron scattering angle.}
\label{deutt20}
\end{figure}

\begin{figure}
\caption{The square of the deuteron magnetic form factor calculated with
(full line) and without (dashed line) convection current term.}
\label{deutmq}
\end{figure}

\begin{figure}
\caption{The transition density $\rho_{tr}(x^{\prime}z^{\prime})$ for
elastic magnetic scattering by deuterons. The peaks are located at
$z^{\prime}$=0 and $x^{\prime}$=$\pm$0.5 fm.}
\label{tranden}
\end{figure}

\begin{figure}
\caption{The deuteron momentum distribution for selected values of $M_d$
and $\theta_k$.}
\label{rhodk}
\end{figure}

\begin{figure}
\caption{The calculated values of $\vec{d}(e,e^{\prime}p)n$ cross section
for the kinematics described in the text. Hollow and filled symbols indicate
results of complete calculations without and with meson-exchange currents.}
\label{2heep}
\end{figure}

\begin{figure}
\caption{$\rho_{0,1}^{M_S}(r,\theta)/R_{Ad}$ for various nuclei.}
\label{rhoA}
\end{figure}

\begin{figure}
\caption{$R_0(r_{dp})$ and $R_2(r_{dp})$ for $^3$He. The points show results
of Monte Carlo calculations in configuration space, and the curves are
smooth fits. The asymptotic $R_L$ given by Eq. (5.5) are shown by dashed
lines.}
\label{R02dp}
\end{figure}

\begin{figure}
\caption{Density distribution of $dp$ clusters in $^3$He with
$M_J=\case{1}{2}$,
$M_d$=0 and $M_p=\case{1}{2}$. The peaks are located at $x_{dp}$=0 and
$z_{dp}\sim\pm 1$ fm.}
\label{3Hemd0}
\end{figure}

\begin{figure}
\caption{Density distribution of $dp$ clusters in $^3$He with
$M_J=\case{1}{2}$,
$M_d$=1 and $M_p=-\case{1}{2}$. The peaks are located at $z_{dp}$=0 and
$x_{dp}\sim\pm 1$ fm}
\label{3Hemd1}
\end{figure}

\begin{figure}
\caption{Momentum distribution of $\vec{d} \vec{p}$ clusters in $^3$He
in $M_J=\case{1}{2}$ state for momenta parallel and transverse to the $z$-axis}
\label{3Hemdk}
\end{figure}

\begin{figure}
\caption{$R_0(r_{dd})$ and $R_2(r_{dd})$ for $^4$He. See Fig. 15
for notation.}
\label{R02dd}
\end{figure}

\begin{figure}
\caption{Density distribution of $\vec{d} \vec{d}$ clusters in $^4$He
in parallel ($\theta$=0) and transverse ($\theta$=$\pi/2$) directions.}
\label{4Hedd}
\end{figure}

\begin{figure}
\caption{Momentum distribution of $\vec{d} \vec{d}$ clusters in $^4$He
in parallel ($\theta$=0) and transverse ($\theta$=$\pi/2$) directions.}
\label{4Heddk}
\end{figure}

\begin{figure}
\caption{$R_0(r_{\alpha d})$ and $R_2(r_{\alpha d})$ for $^6$Li. See
Fig. 15 for notation.}
\label{R02ad}
\end{figure}

\begin{figure}
\caption{Density distribution of $\alpha \vec{d}$ clusters in $^6$Li
in parallel ($\theta$=0) and transverse ($\theta$=$\pi/2$) directions.}
\label{6Liad}
\end{figure}

\begin{figure}
\caption{$\rho_{1,1}^{M_S}(r,\theta)/R^{A}_{1,1}$ for various nuclei.
The upper three curves are for $M_S$=0, $\theta$=0 while the lower ones
are for $M_S$=0, $\theta$=$\pi/2$ and equivalently $M_S$=$\pm$1, $\theta$=0.}
\label{ts11fig}
\end{figure}

\begin{figure}
\caption{$\rho_{1,0}^{0}(r)/R^{A}_{1,0}$ for various nuclei. The $\rho(r)$
of an unpolarized deuteron, normalized to have the same maximum value, is
shown for comparison by the dotted line.}
\label{ts10fig}
\end{figure}

\begin{figure}
\caption{$\rho_{0,0}^{0}(r)/R^{A}_{0,0}$ for various nuclei.}
\label{ts00fig}
\end{figure}

\newpage
\begin{table}[h]
\caption{The calculated values of $R_{Ad}$ and other ratios.}
\vspace{0.3in}
\begin{tabular}{cdddddd}
$\  \  $Nucleus    &      $R_{Ad}$    &
   $\frac{\langle v_{\pi}\rangle_A}{\langle v_{\pi}\rangle_d}$      &
   $\frac{\sigma_{ab,A}^{\pi}}{\sigma_{ab,d}^{\pi}}$ &
   $\frac{\sigma_{ab,A}^{\gamma}}{\sigma_{ab,d}^{\gamma}}$ &
   \multicolumn{2}{c} {$N^{A}_{0,1}$} \\
  &  &  &  &  & IP & $\Psi_v$ \\
\tableline
$^3$He    &   2.0    &    2.1    &   2.4(1) & $\sim$2 & 1.5 & 1.49  \\
$^4$He    &   4.7    &    5.1    &   4.3(6) & $\sim$4 & 3.  & 2.99  \\
$^6$Li    &   6.3    &    6.3    &          &         & 5.5 & 5.46  \\
$^7$Li    &   7.2    &    7.8    &          & 6.5(5)  & 6.75 & 6.73      \\
$^{16}$O  &  18.8    &   22      &  17(3)   & 16(3)   & 30. & 30.1  \\
\end{tabular}

\vspace{1.0in}

\caption{The calculated values of $R_{T,S}^A$ and $N_{T,S}^A$ in various
nuclei.}
\vspace{0.3in}
\begin{tabular}{cddddddddd}
$\  \  $Nucleus    &
     $R^A_{1,0}$  & \multicolumn{2}{c} {$N^{A}_{1,0}$} &
     $R^A_{0,0}$  & \multicolumn{2}{c} {$N^{A}_{0,0}$} &
     $R^A_{1,1}$  & \multicolumn{2}{c} {$N^{A}_{1,1}$} \\
  &  & IP & \multicolumn{1}{c} {$\Psi_v$} &
     & IP & \multicolumn{1}{c} {$\Psi_v$} &
     & IP & \multicolumn{1}{c} {$\Psi_v$} \\
\tableline
$^3$He    & 0.087 & 1.5  & 1.35 & 0.0016 & 0.   & 0.01 & 0.012 & 0.   & 0.14 \\
$^4$He    & 0.22  & 3.   & 2.5  & 0.0085 & 0.   & 0.01 & 0.060 & 0.   & 0.47 \\
$^6$Li    & 0.24  & 4.5  & 4.0  & 0.061  & 0.5  & 0.52 & 0.104 & 4.5  & 4.96 \\
$^7$Li    & 0.37  & 6.75 & 6.1  & 0.118  & 0.75 & 0.77 & 0.18  & 6.75 & 7.41 \\
$^{16}$O  & 1.    & 30.  & 28.5 & 1.     & 6.   & 6.05 & 1.    & 54.  & 55.5 \\
\end{tabular}
\end{table}

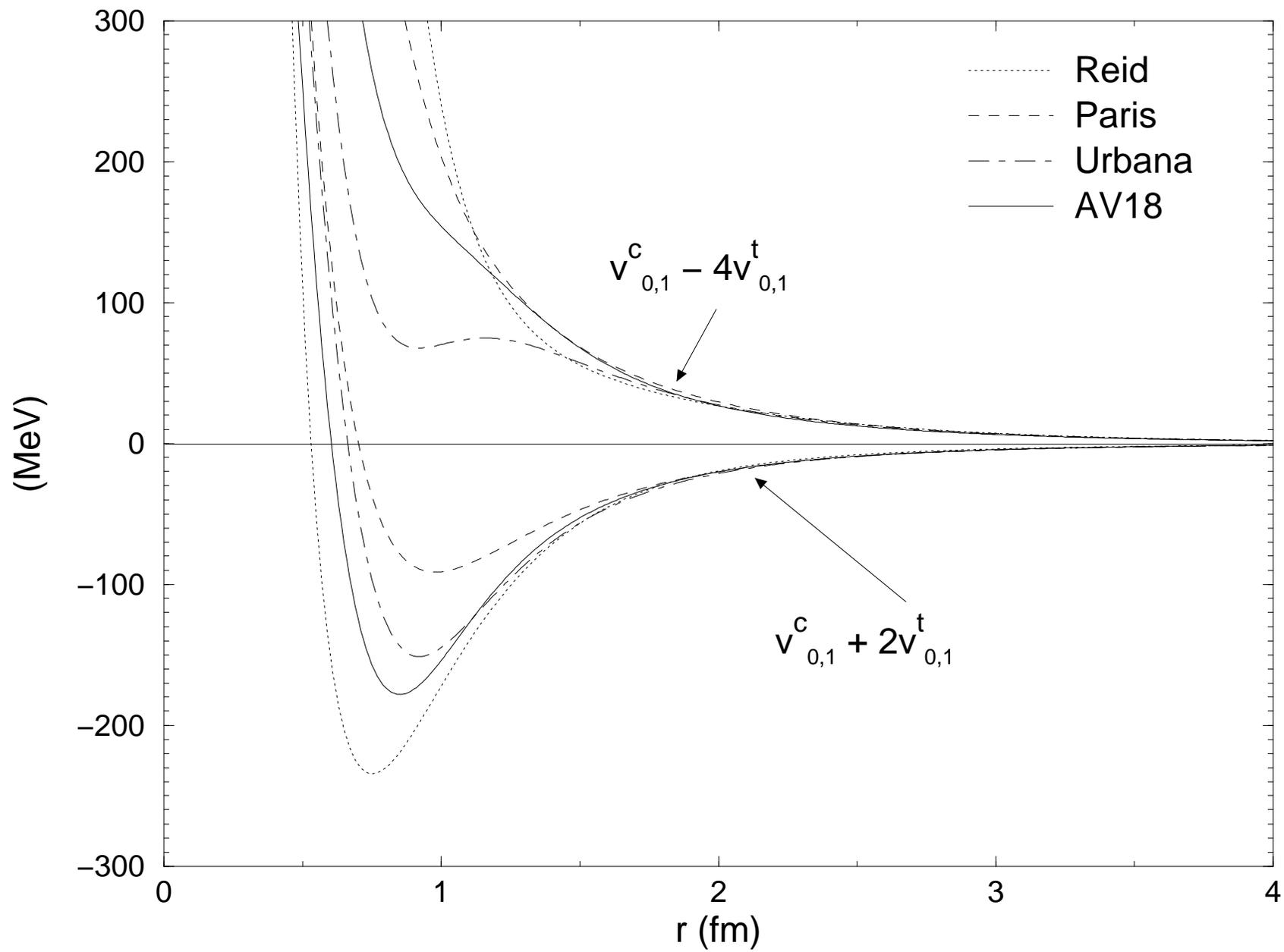
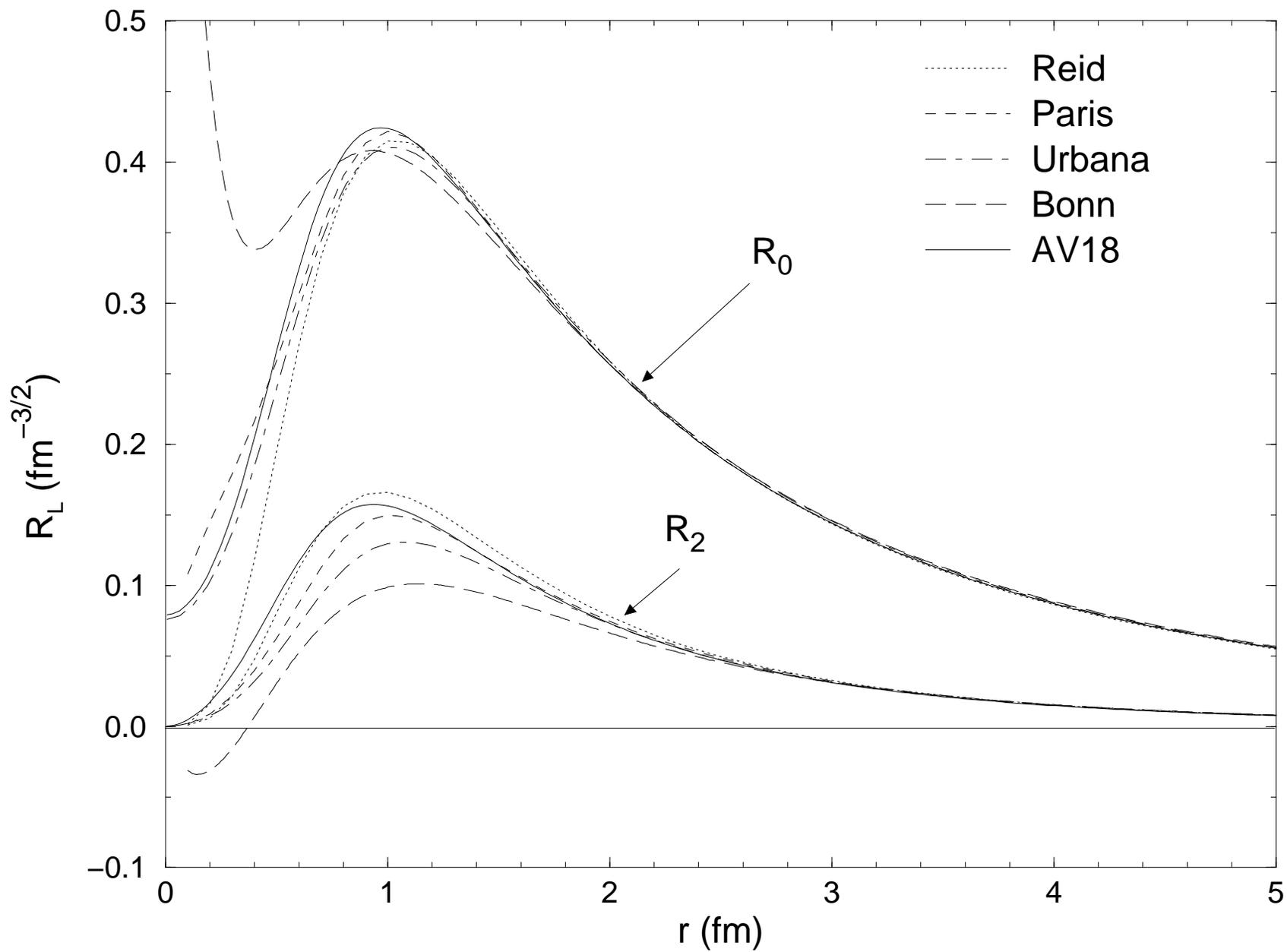
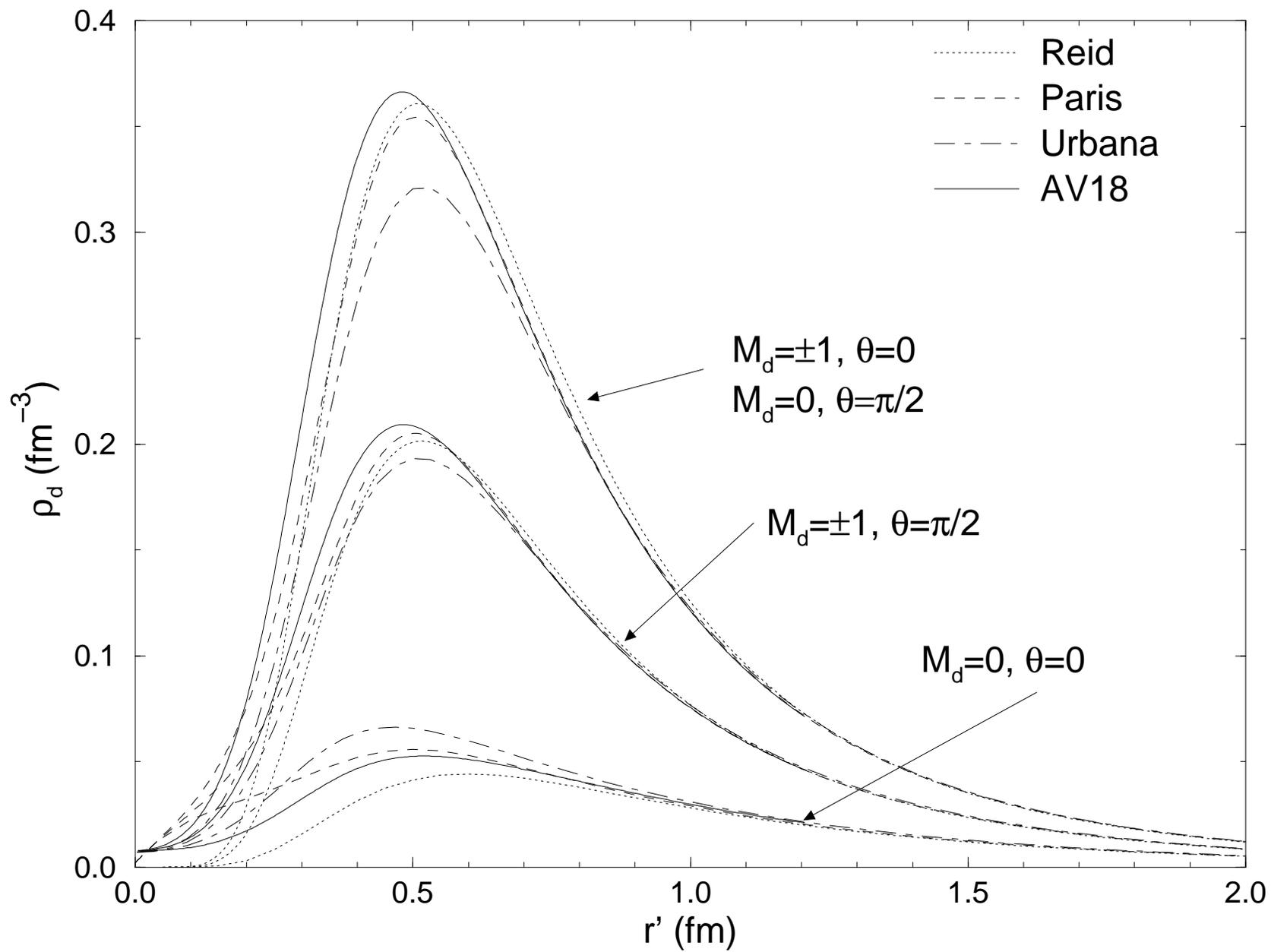
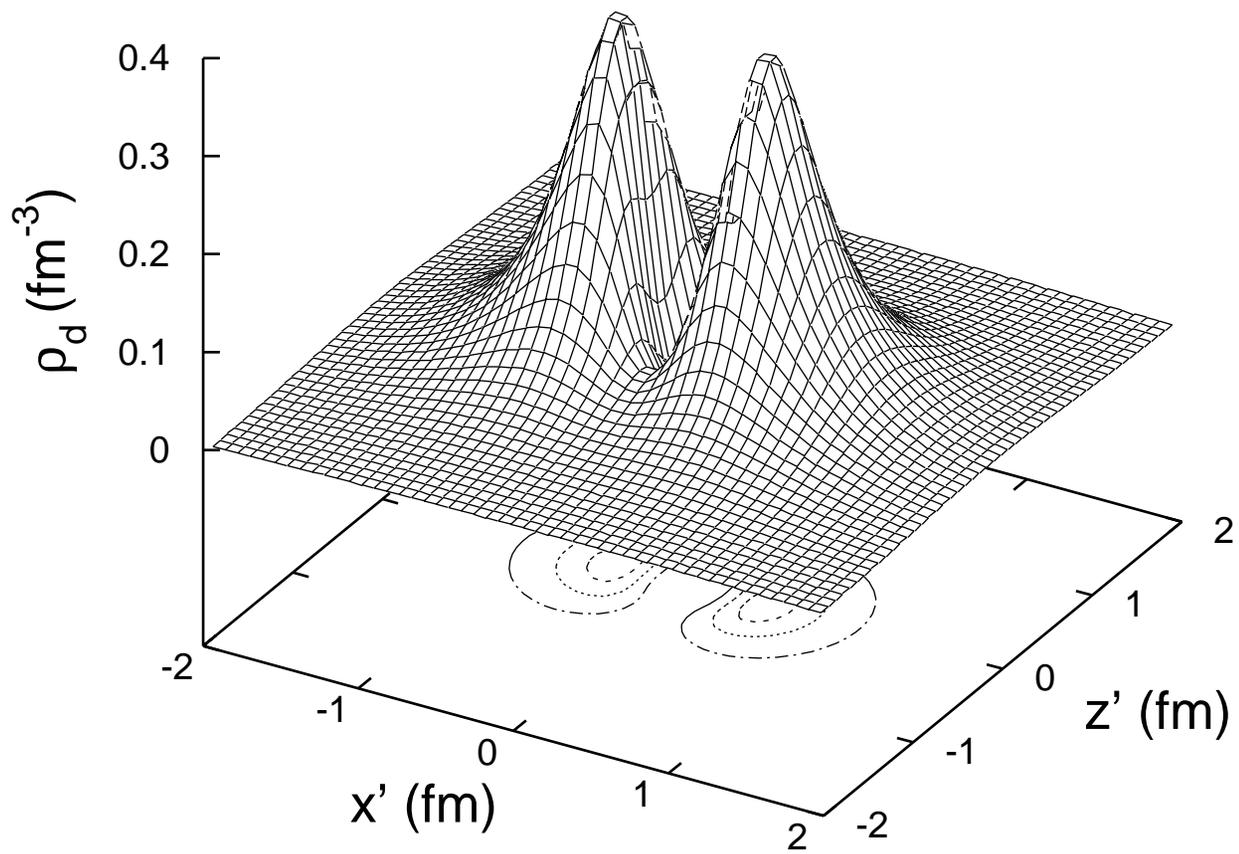
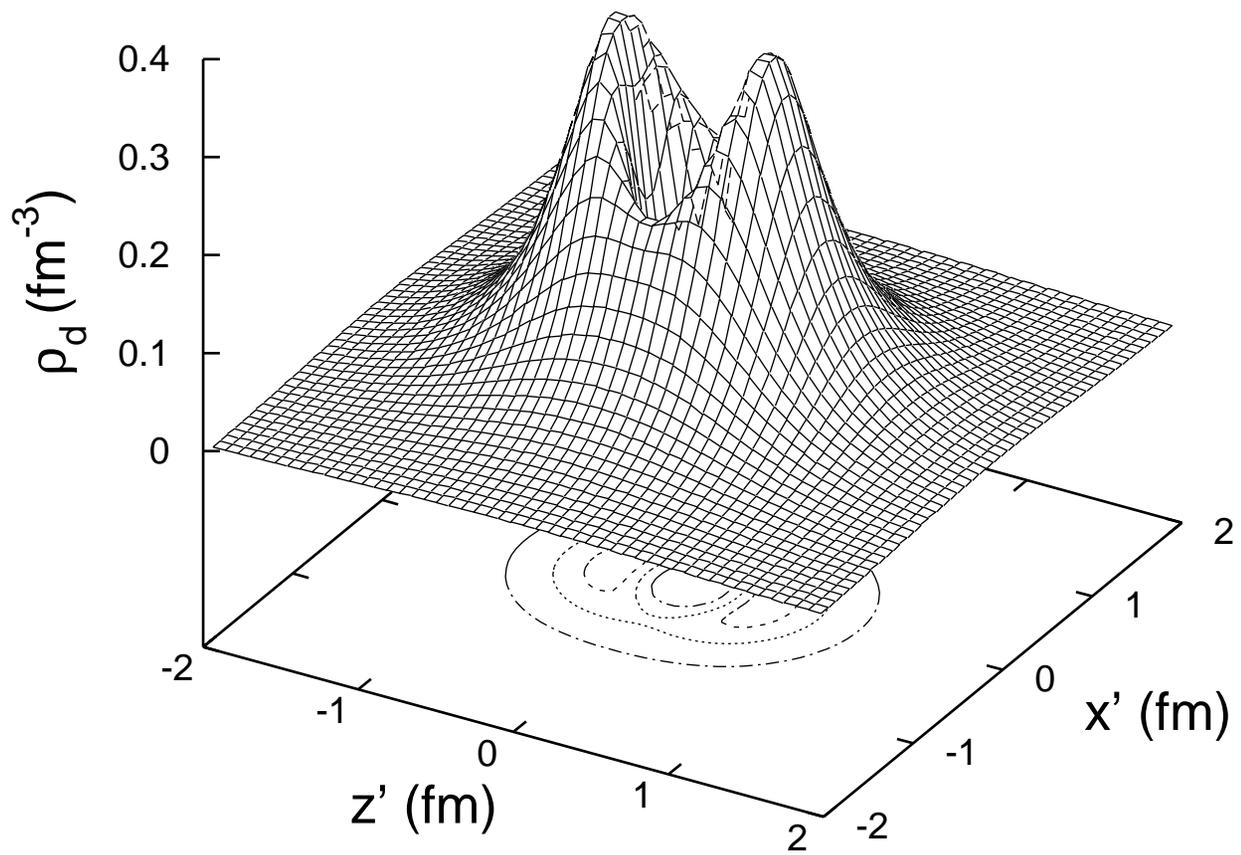
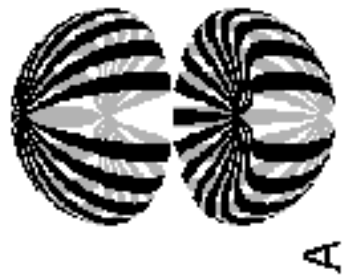
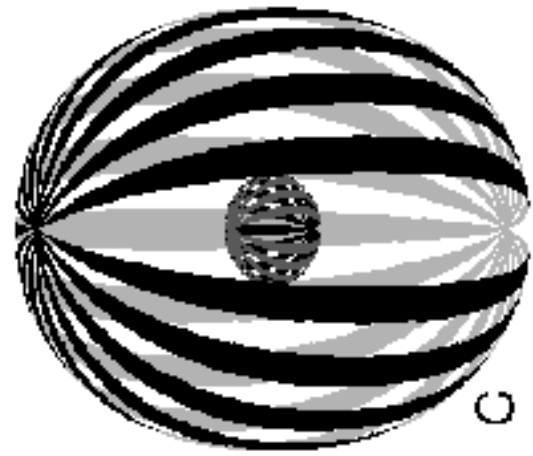
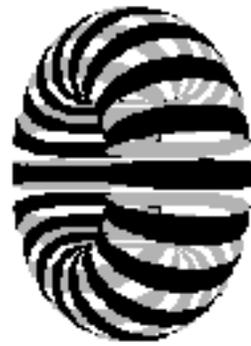
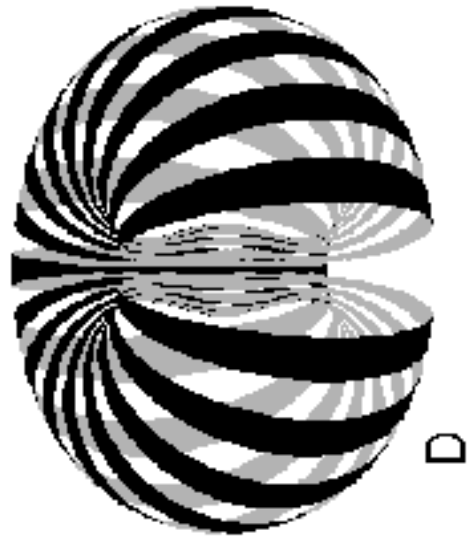
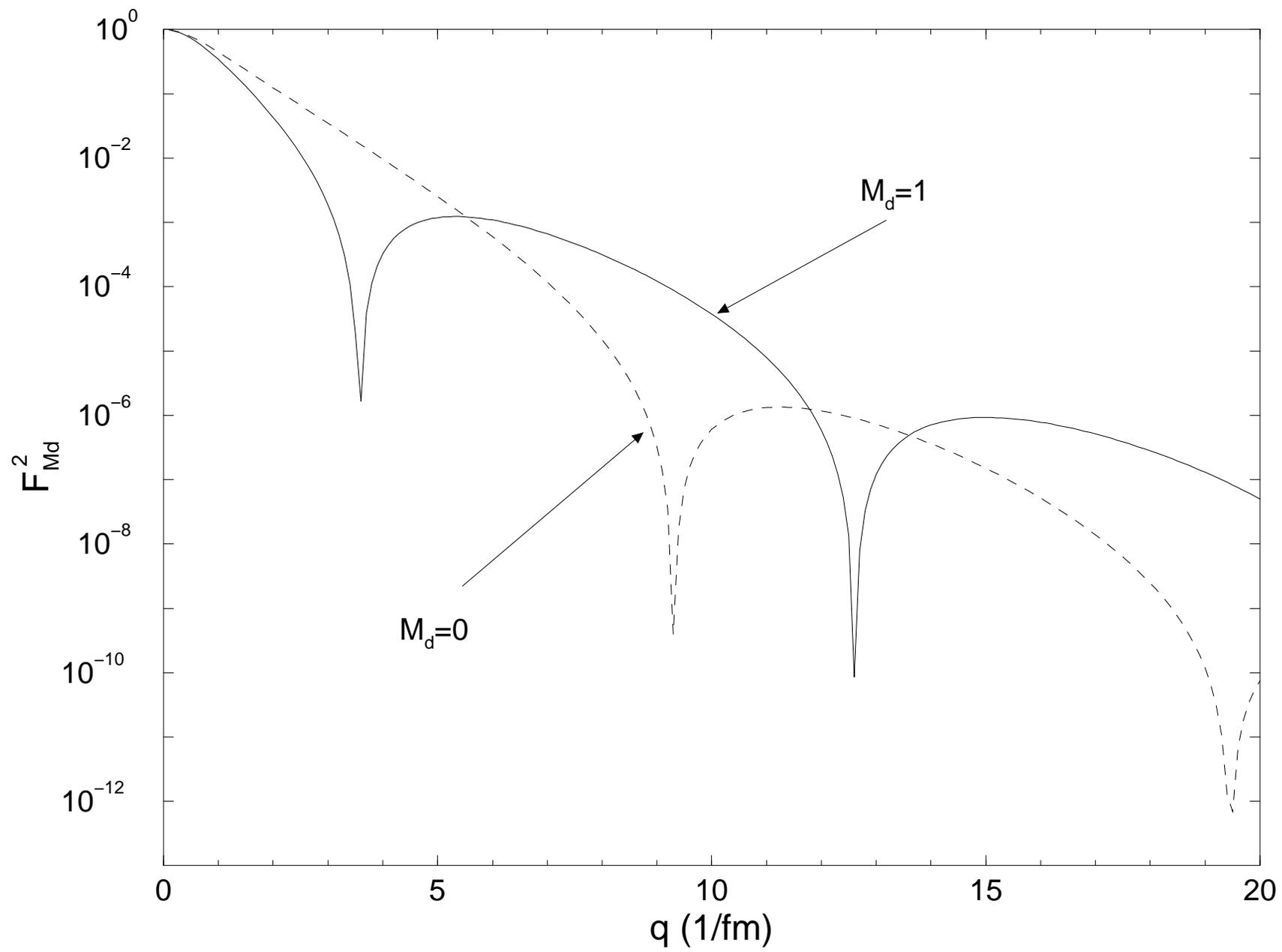
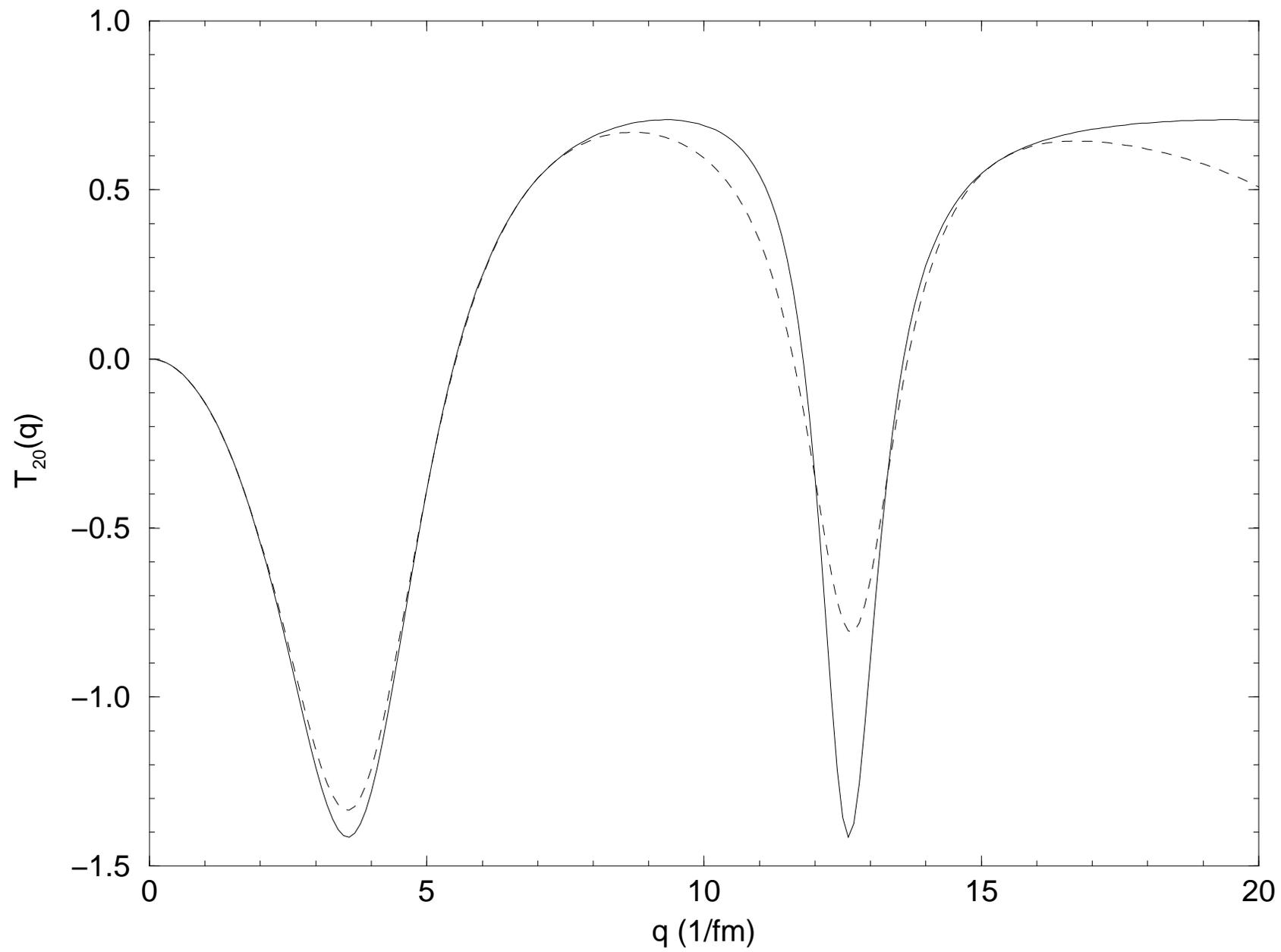
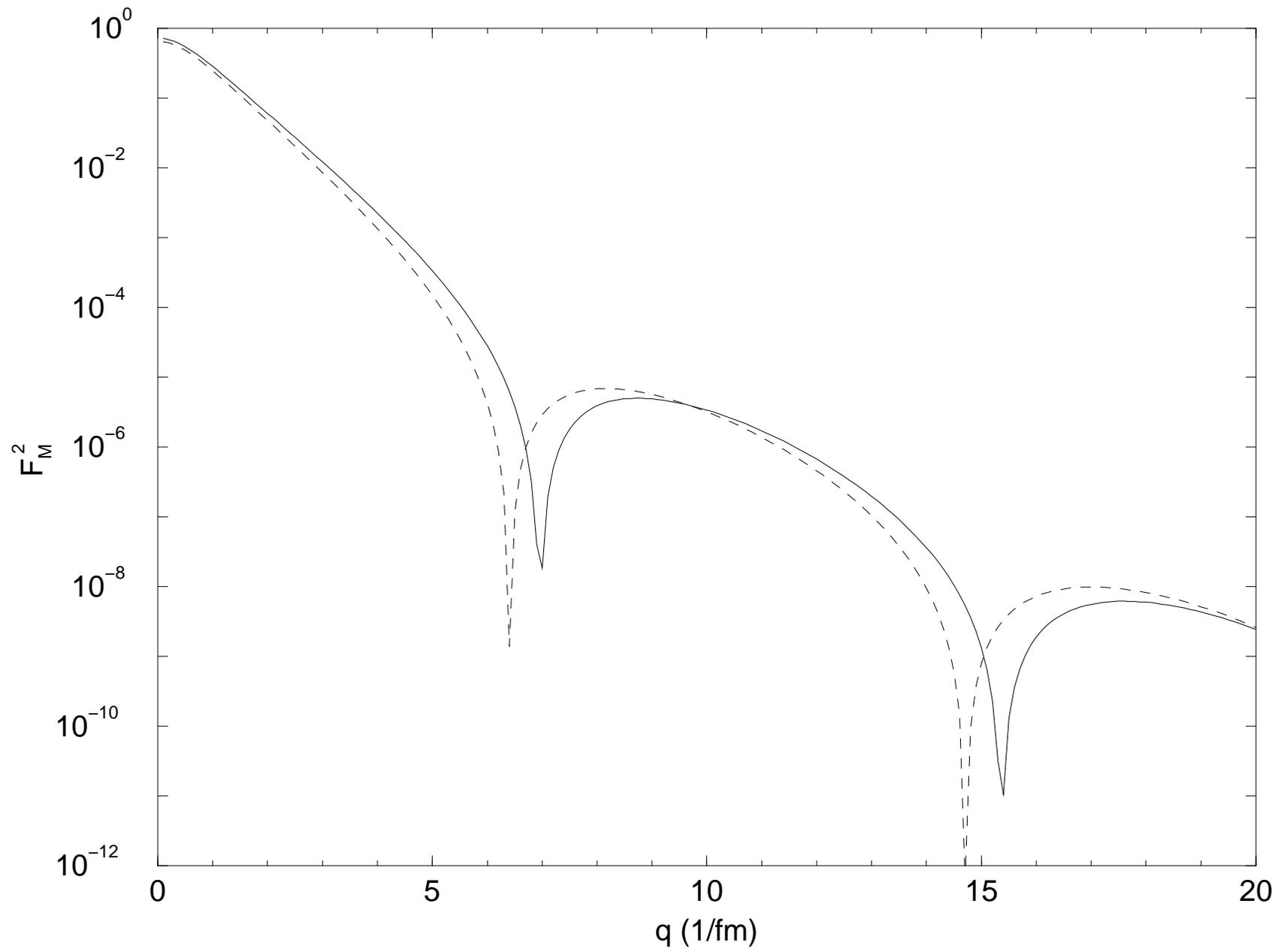
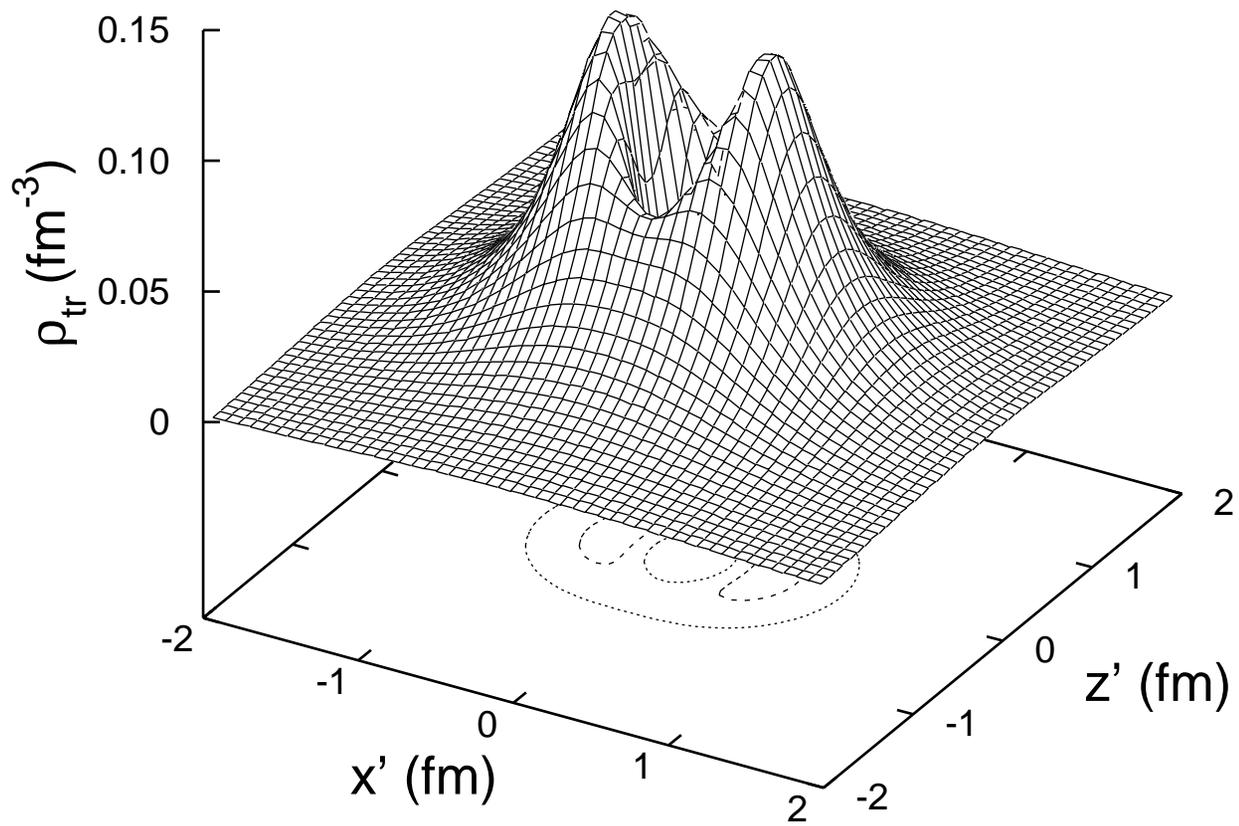
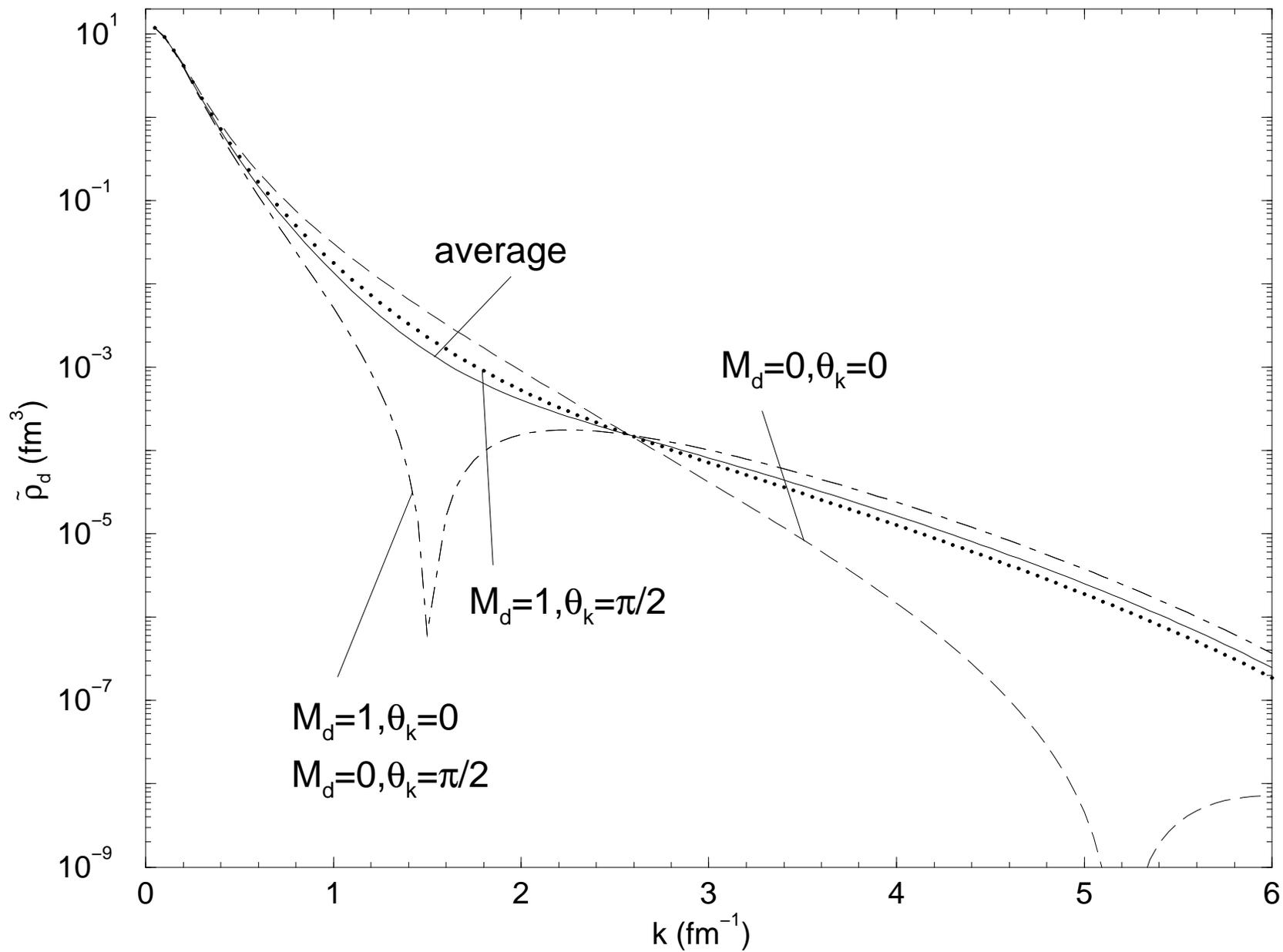
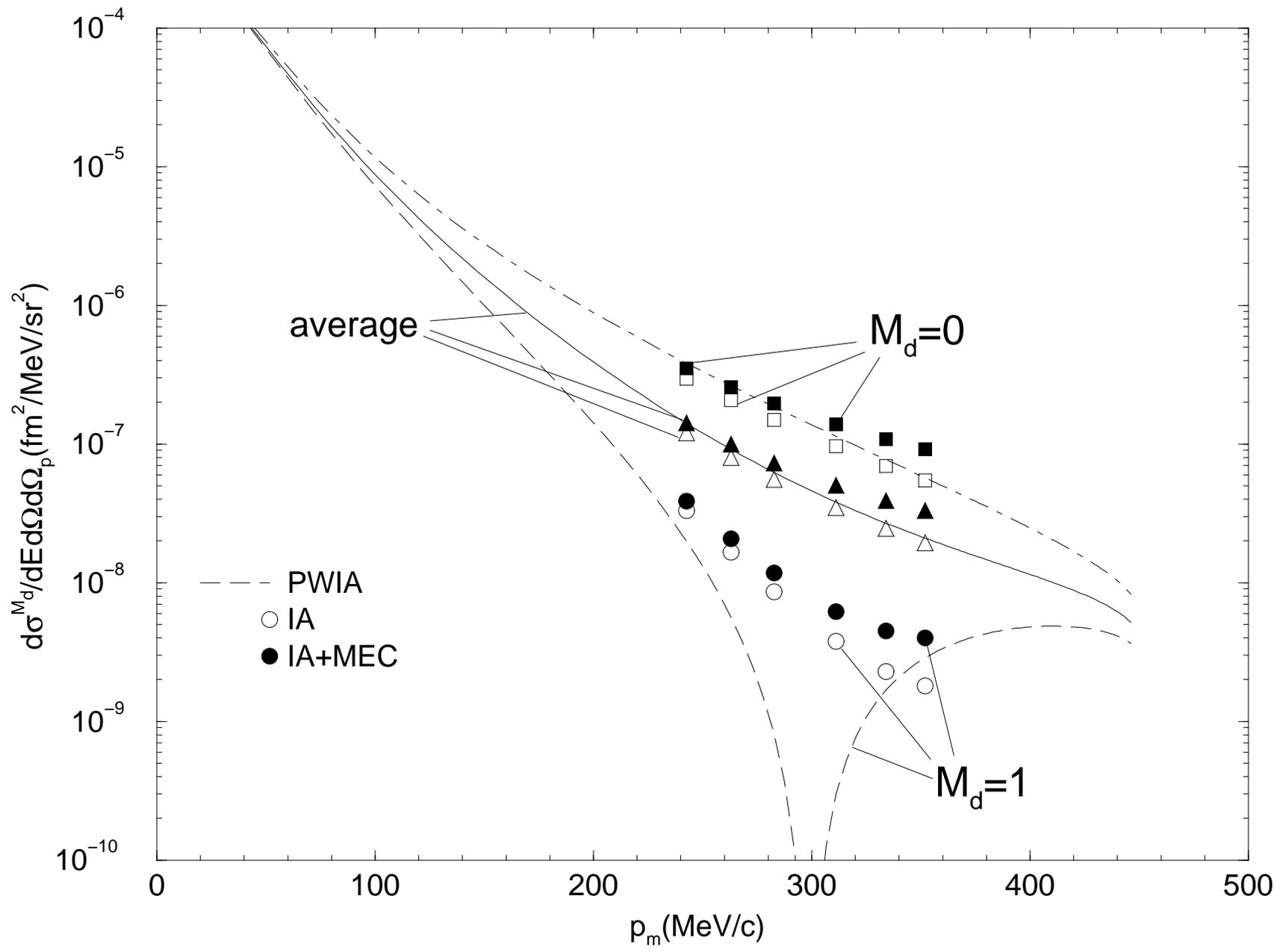
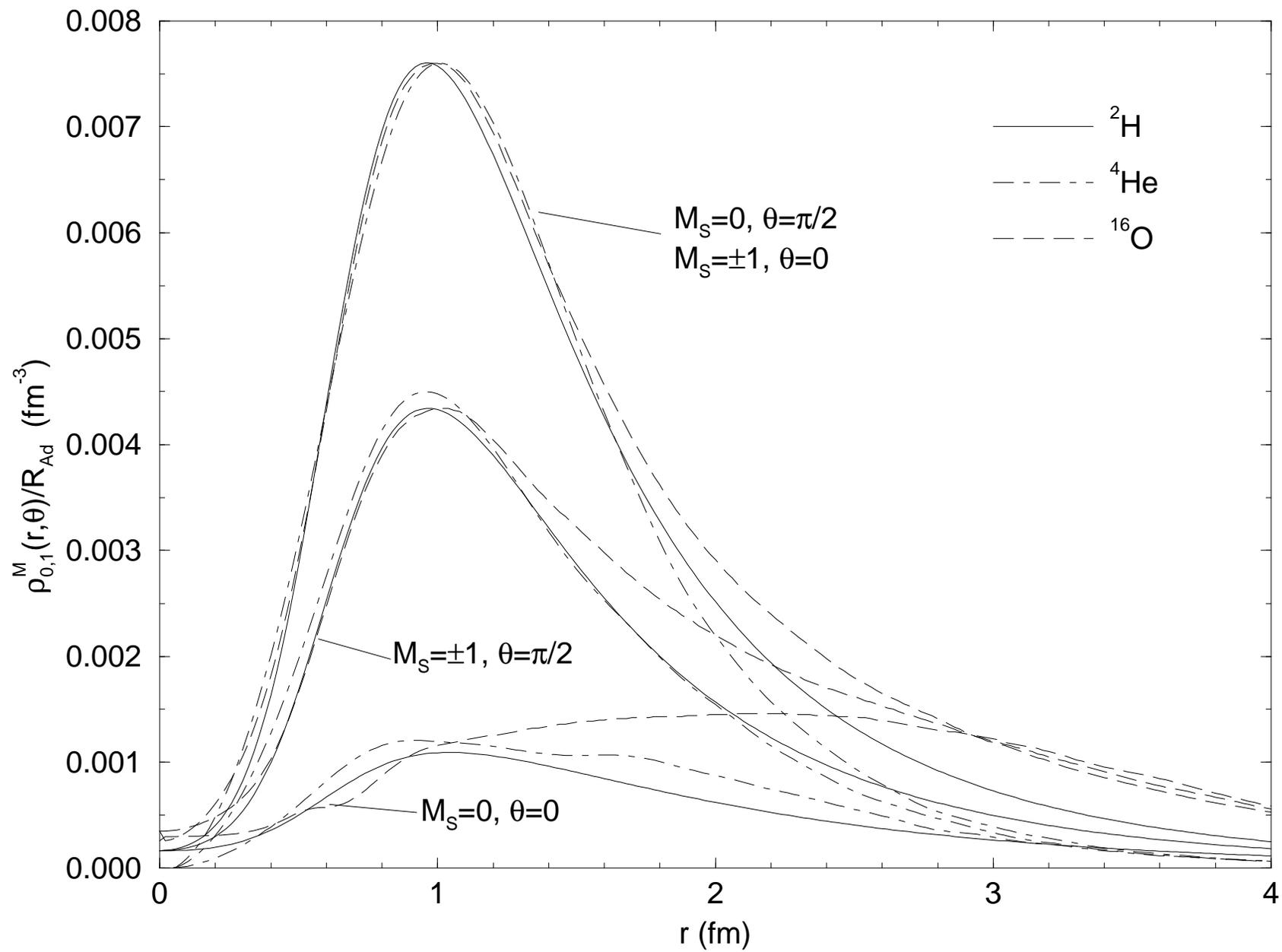
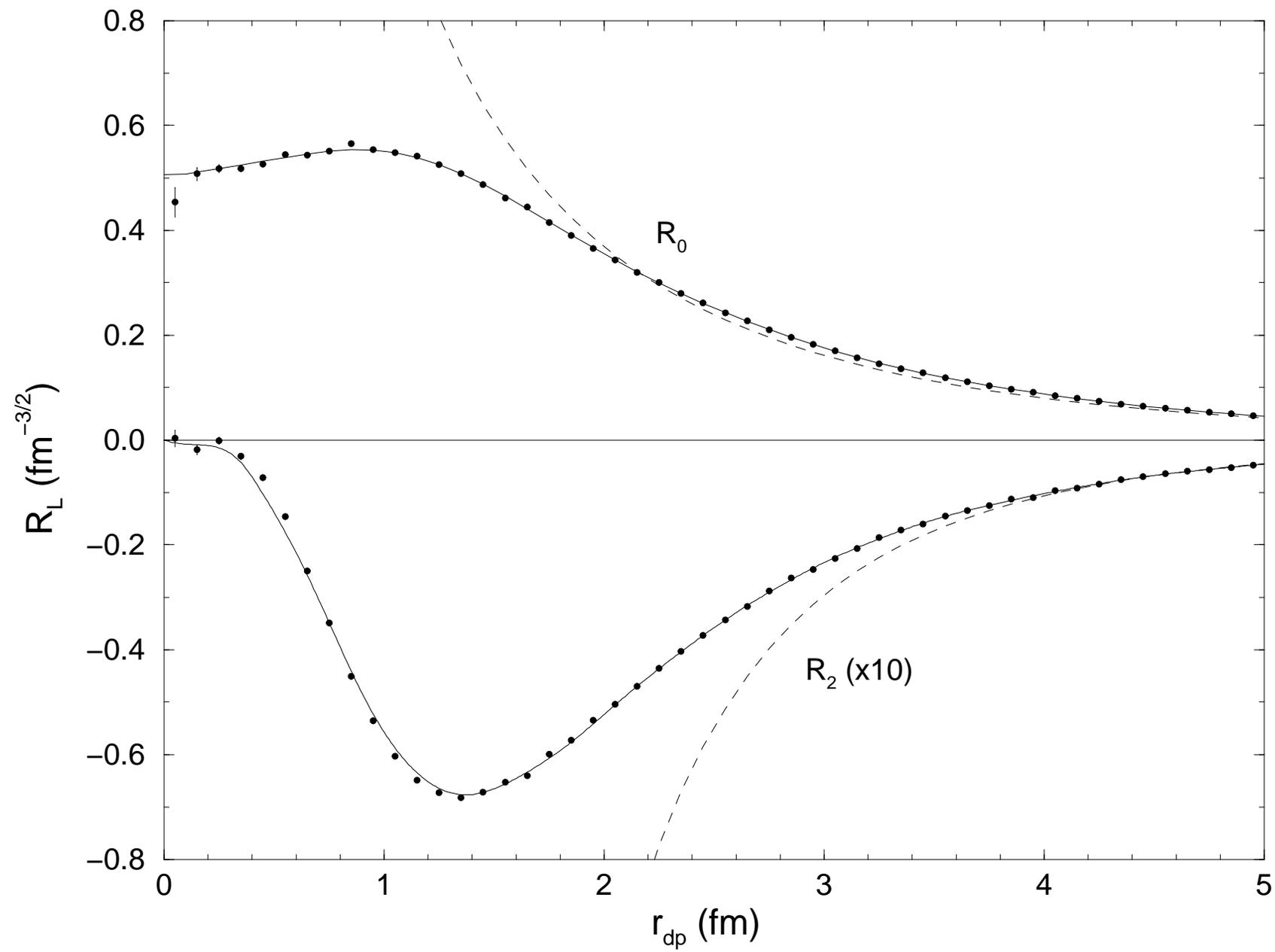
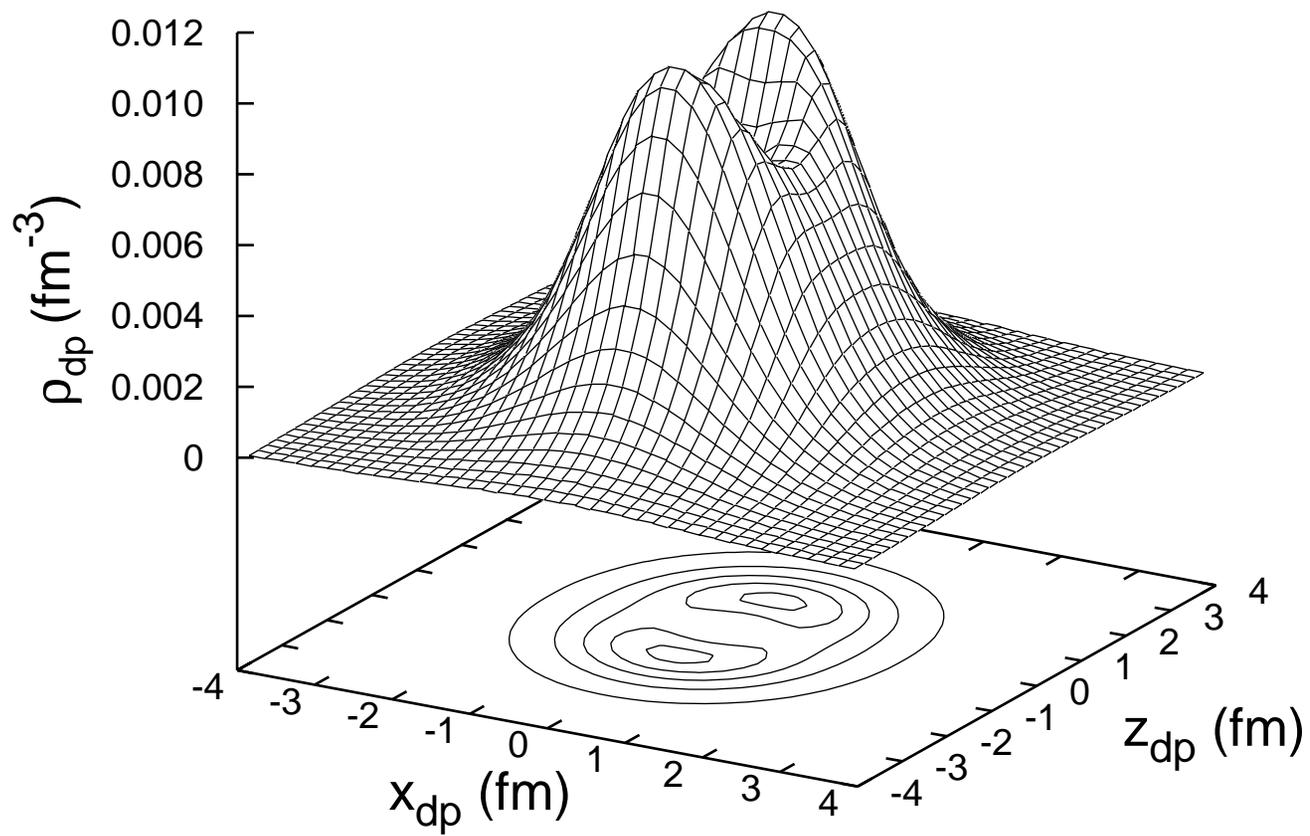
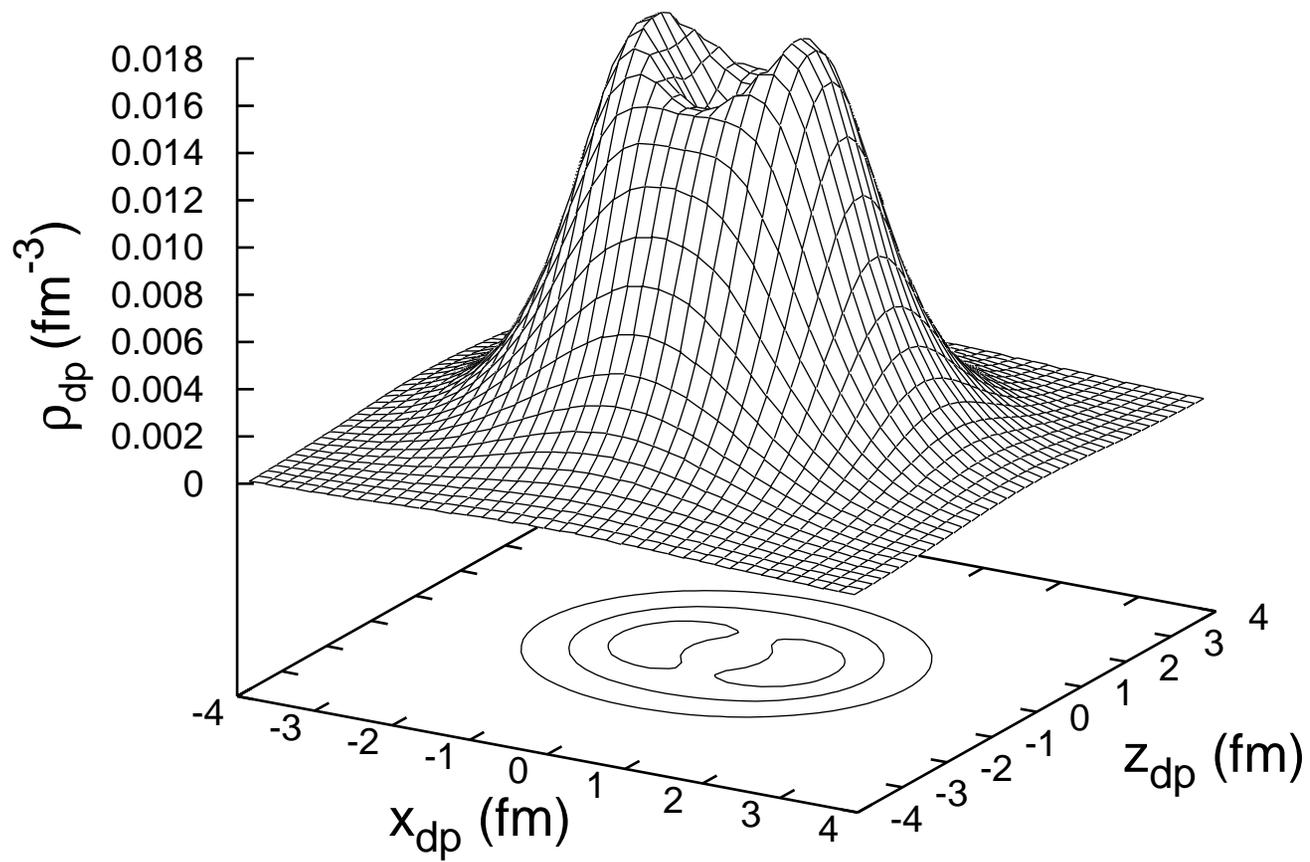
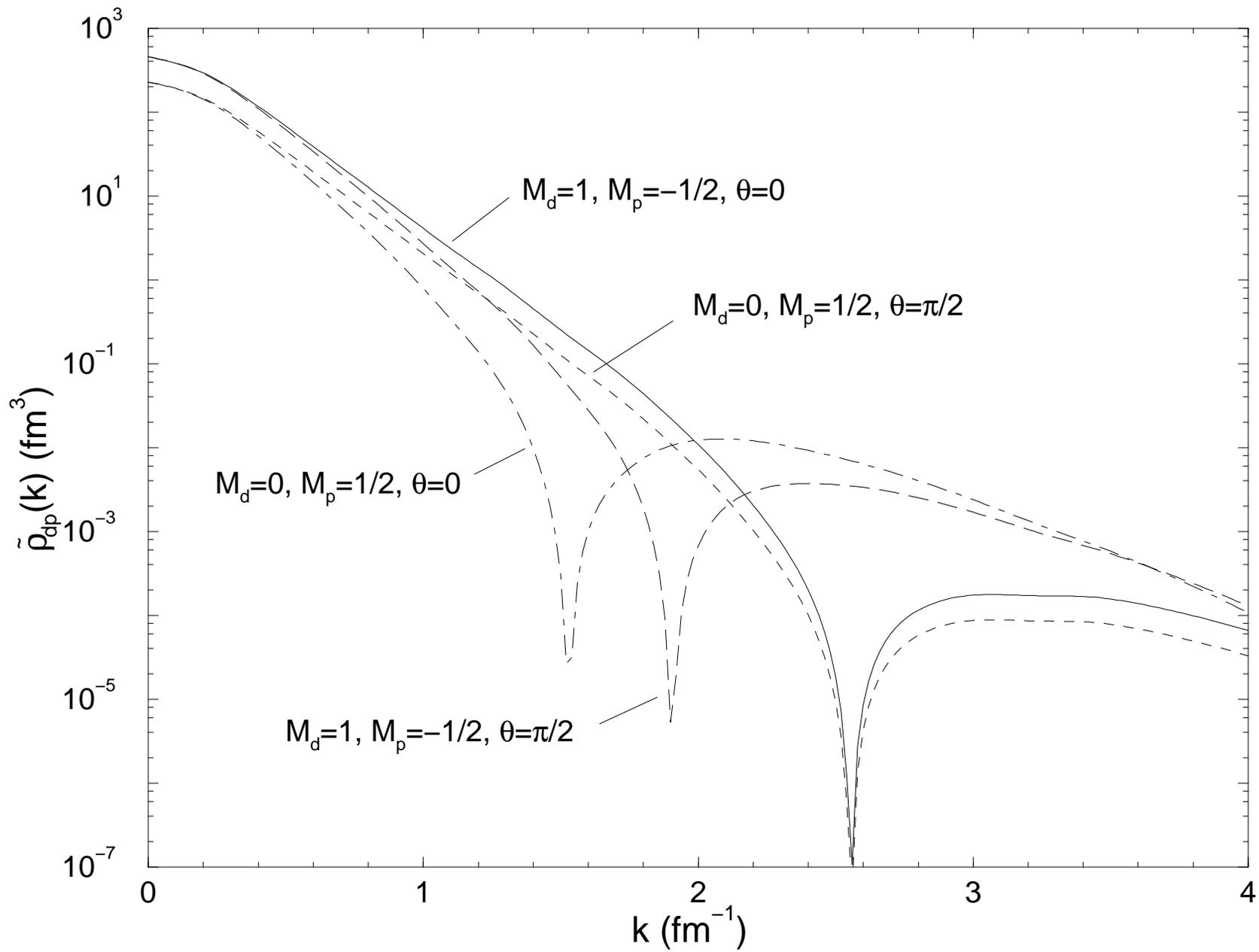
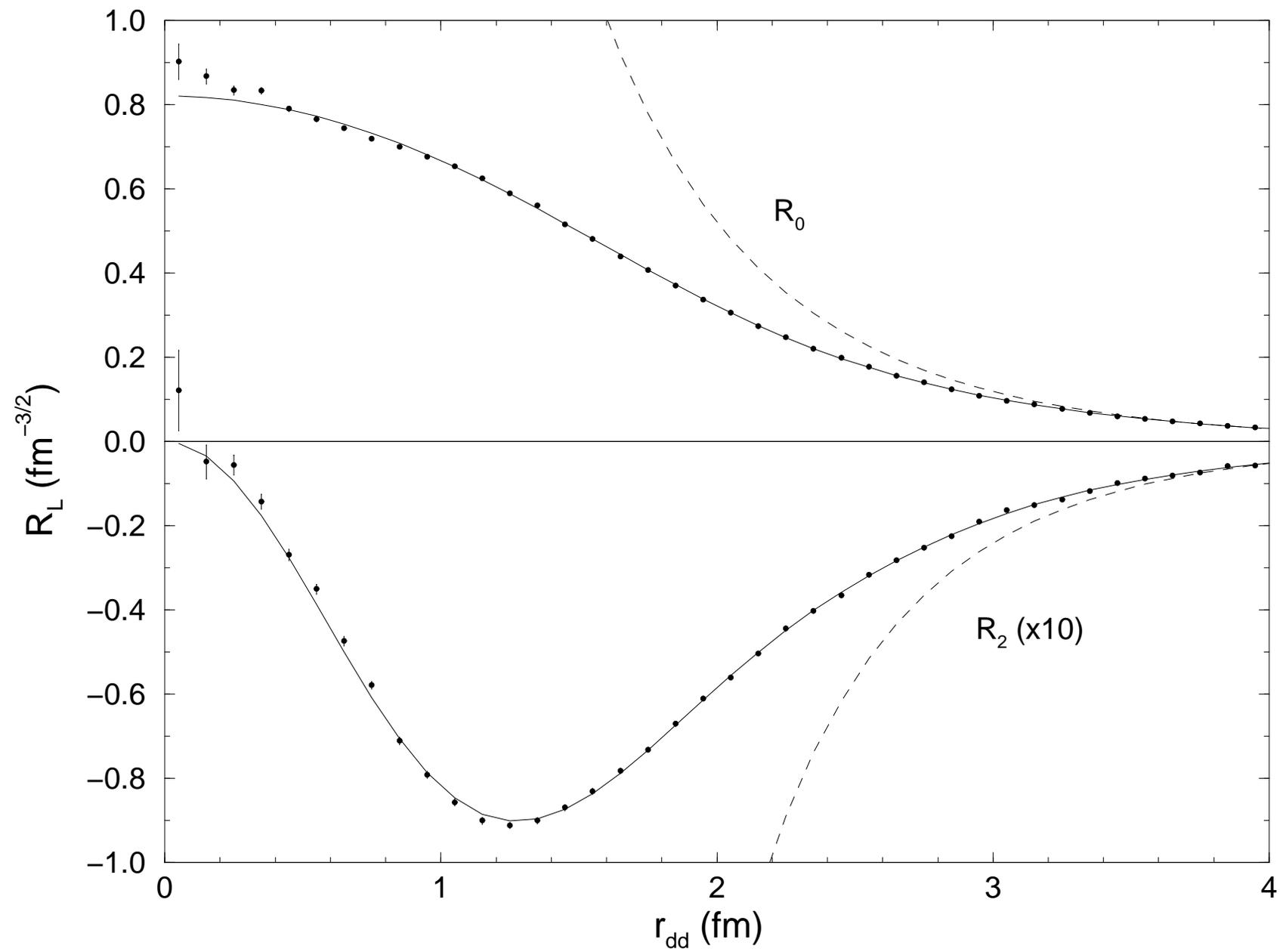
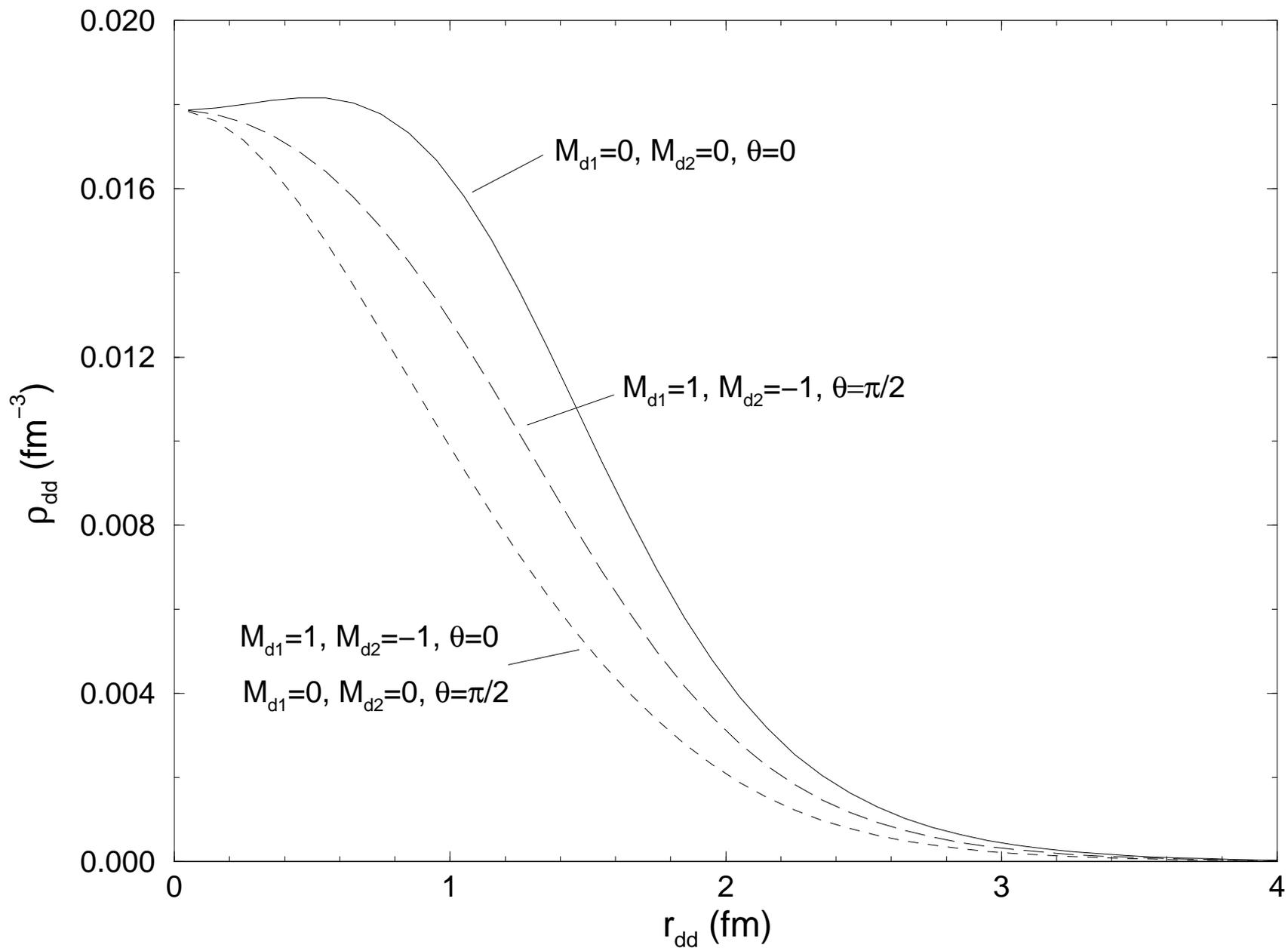
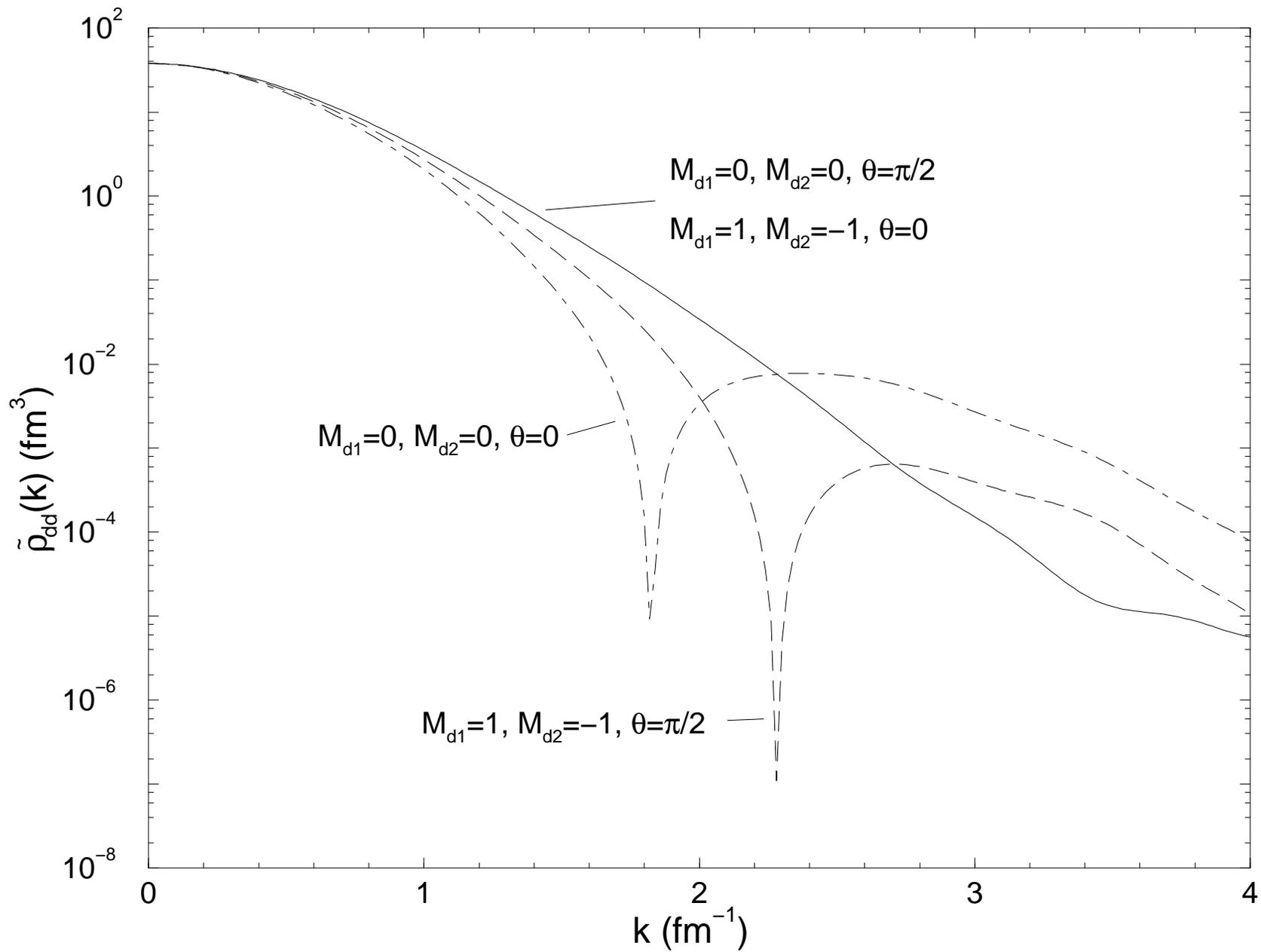
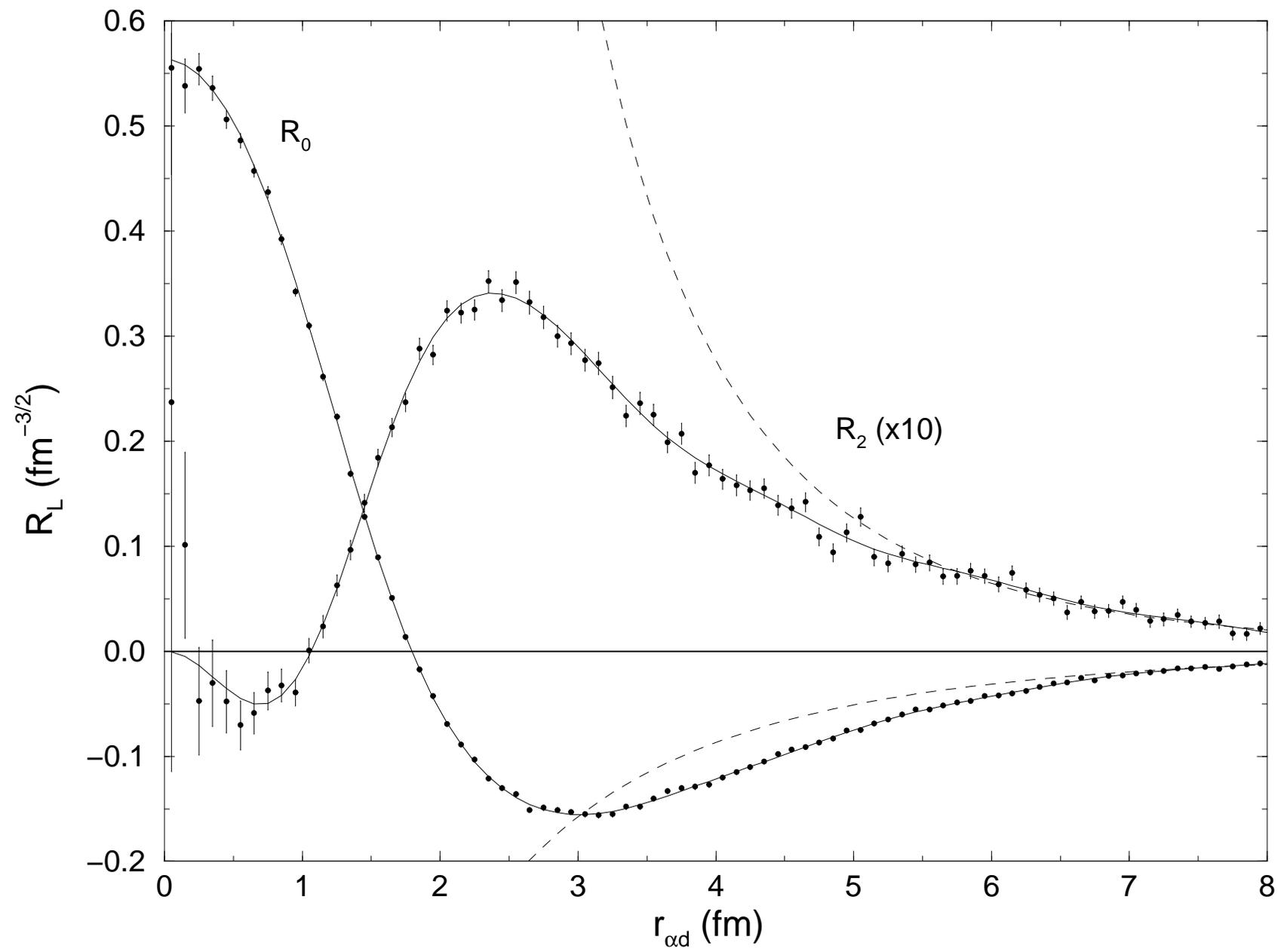
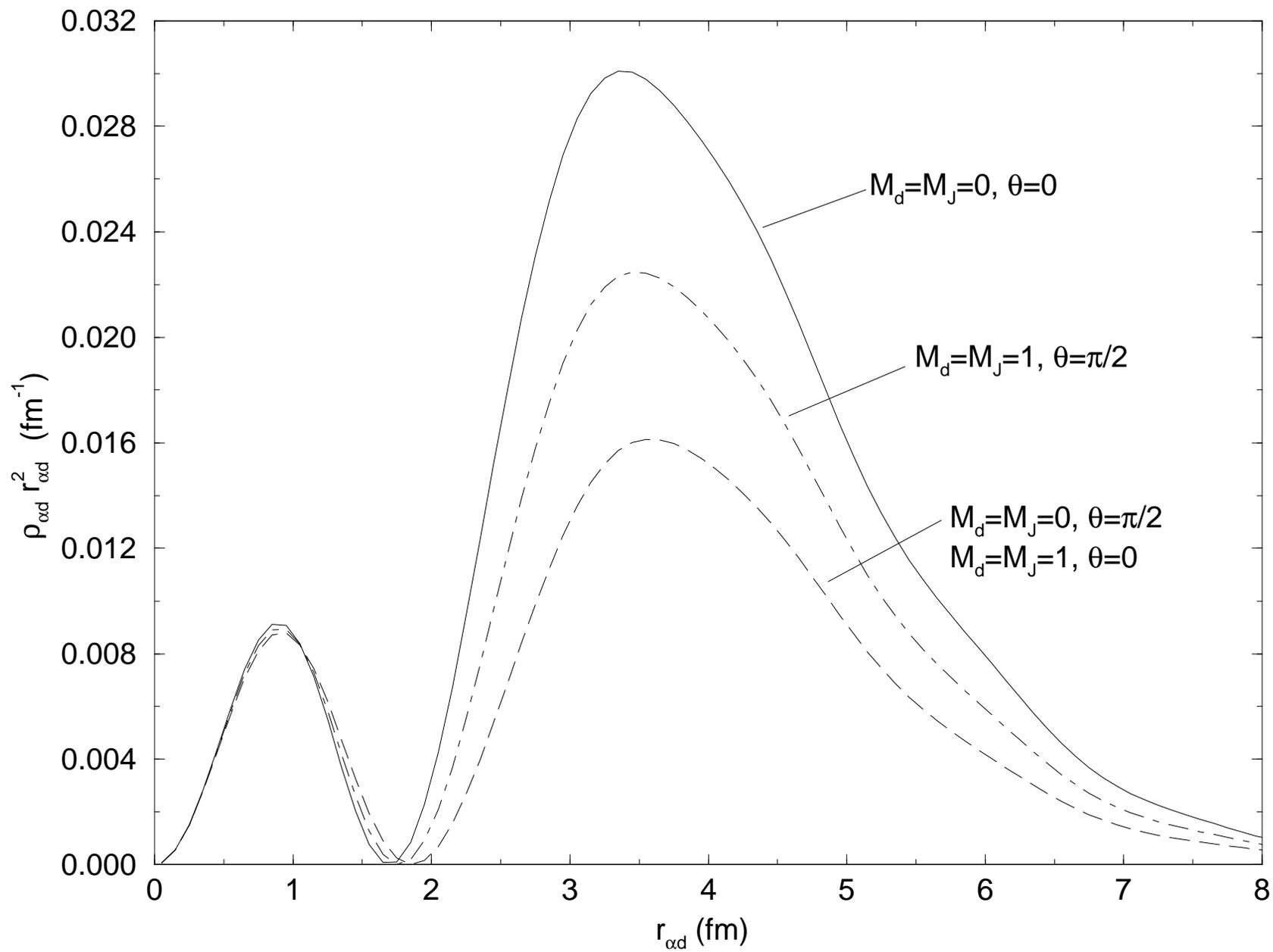
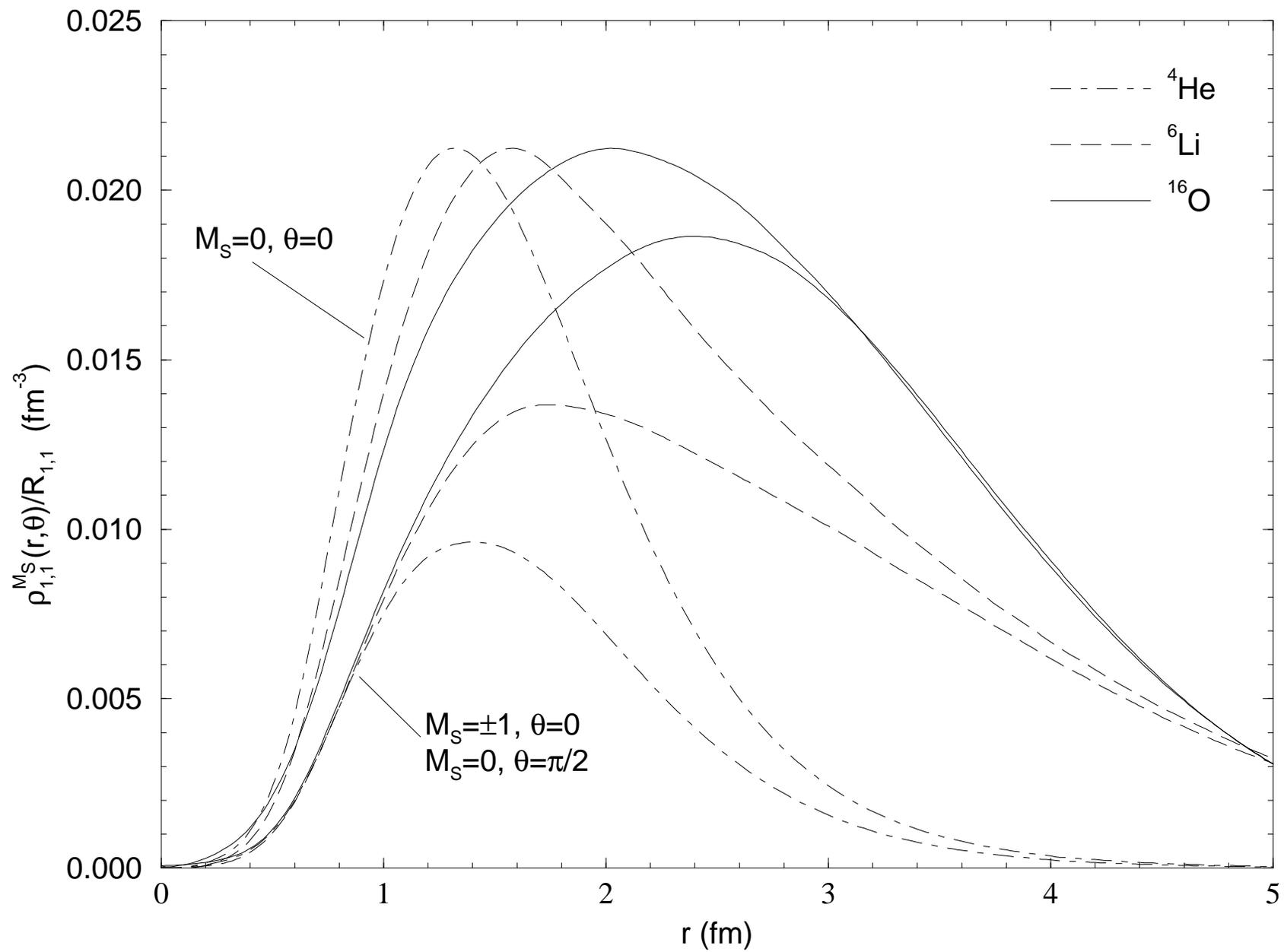
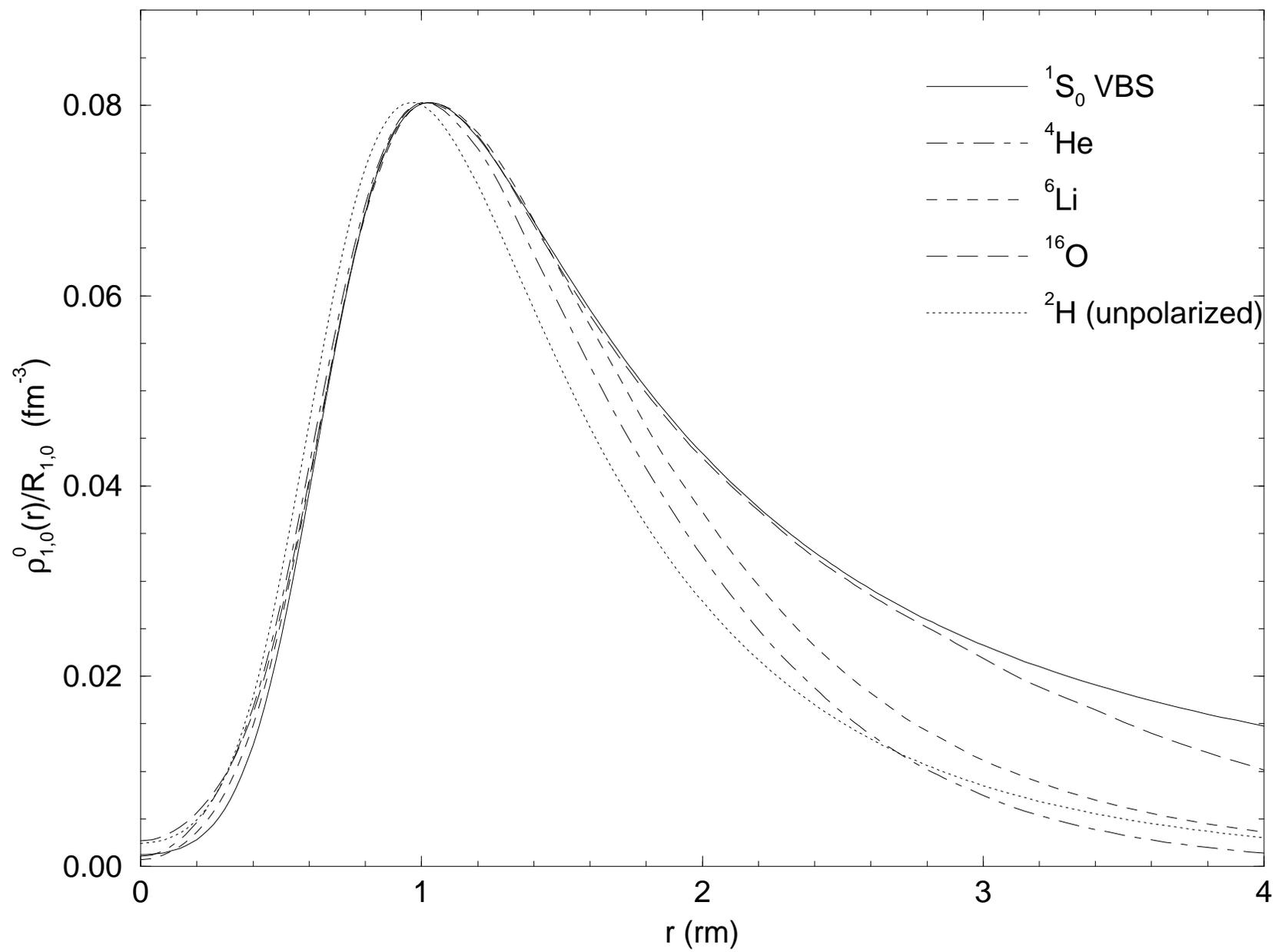
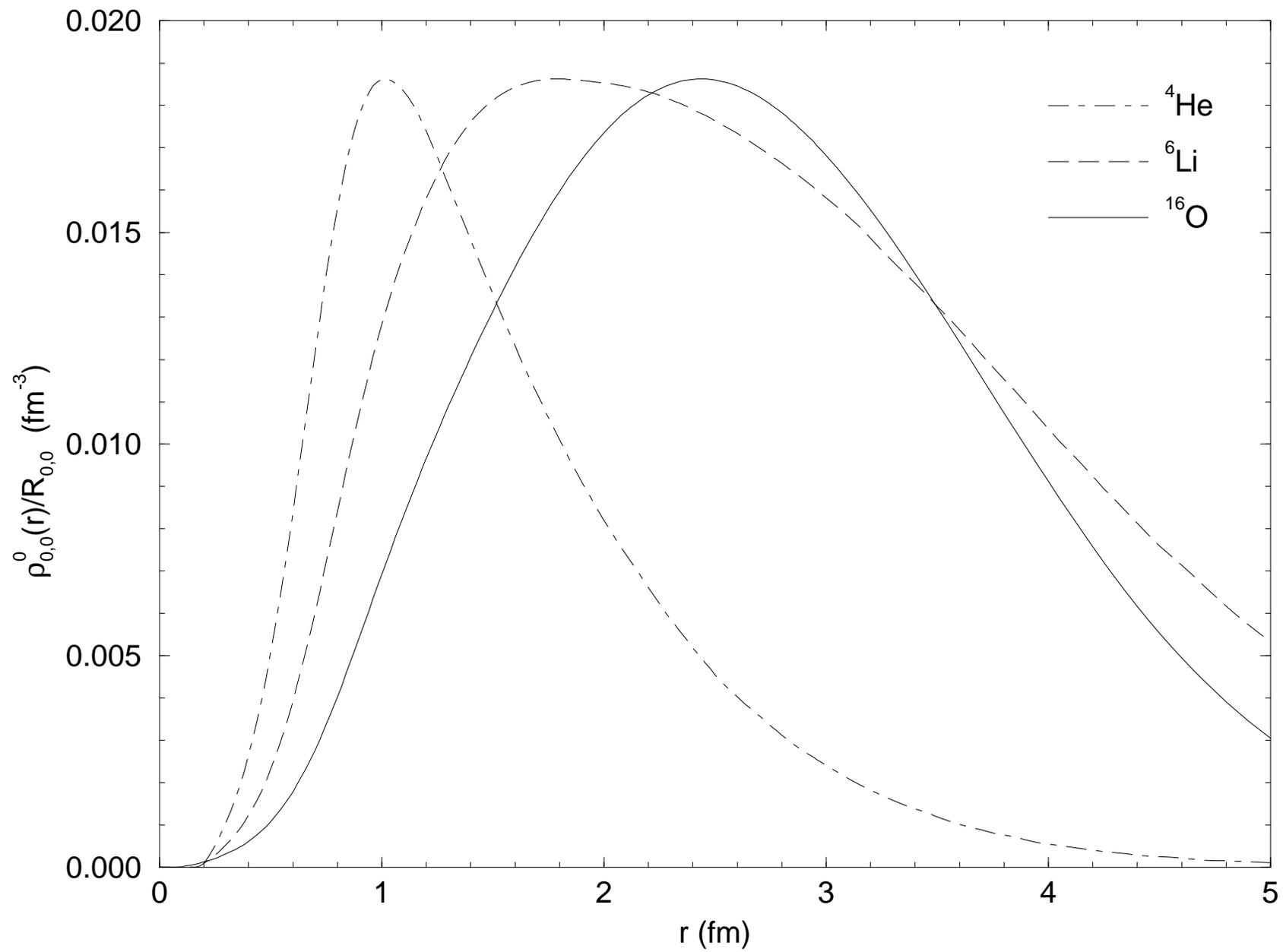
\end{document}